\documentclass[12pt,letterpaper]{article}
\usepackage[english]{babel}
\usepackage[utf8]{inputenc}
\usepackage[a4paper, total={6. in, 8in}]{geometry}
\usepackage[colorlinks, linkcolor=black,citecolor=black, urlcolor=black,bookmarks=true]{hyperref} 
\usepackage{csvsimple} 
\usepackage{graphicx}
\usepackage{dcolumn}
\usepackage{bm}
\usepackage{amsmath}
\usepackage[utf8]{inputenc}
\usepackage[T1]{fontenc}
\usepackage{mathptmx}
\usepackage{etoolbox}
\usepackage{xr}
\usepackage{cite}
\usepackage{longtable}

\author{Flaviano Della Pia $^{1}$, Andrea Zen $^{2,3}$, Dario Alfè $^{2,3,4,5}$, and Angelos Michaelides $^{1}$ \\\\
        \small $^{1}$Yusuf Hamied Department of Chemistry, University of Cambridge, Cambridge CB2 1EW, United Kingdom \\
        \small $^{2}$ Dipartimento di Fisica Ettore Pancini, Università di Napoli Federico II, Monte S. Angelo, I-80126 Napoli, Italy \\
        \small $^{3}$ Department of Earth Sciences, University College London, London WC1E 6BT, United Kingdom \\
        \small $^{4}$ Thomas Young Centre, University College London, London WC1E 6BT, United Kingdom\\
        \small $^{5}$ London Centre for Nanotechnology, University College London, London WC1E 6BT, United Kingdom \\
}

\title{
How accurate are simulations and experiments for the lattice energies of molecular crystals?
}
\date{}
\begin{document}
\maketitle

\begin{abstract}
Molecular crystals play a central role in a wide range of scientific fields, including pharmaceuticals and organic semiconductor devices. 
However, they are challenging systems to model accurately with computational approaches because of a delicate interplay of intermolecular interactions such as hydrogen bonding and van der Waals dispersion forces.
Here, by exploiting recent algorithmic developments, we report the first set of diffusion Monte Carlo lattice energies for all 23 molecular crystals in the popular and widely used X23 dataset. 
Comparisons with previous state-of-the-art lattice energy predictions (on a subset of the dataset) and a careful analysis of experimental sublimation enthalpies reveals that high-accuracy computational methods are now at least as reliable as (computationally derived) experiments for the lattice energies of molecular crystals. 
Overall, this work demonstrates the feasibility of high-level explicitly correlated electronic structure methods for broad benchmarking studies in complex condensed phase systems, and signposts a route towards closer agreement between experiment and simulation. 
\end{abstract}

\section{Introduction}\label{sec:Introduction}


\noindent Molecular crystals are of central importance to pharmaceuticals\cite{CSD-pharma}, organic semiconductor devices\cite{Mei-molcrys-FET, Corminboeuf-molcrys-organic-semiconductors}, optoelectronics \cite{Ostroverkhova-molcrys-optoelectronics}, and medicine\cite{Grant-molcrys-medicine}. 
Computational approaches play a central role in molecular crystal research, both in aiding experimental structure determination and in predicting their stability.
%
In particular, the computation of lattice energies is pivotal in Crystal Structure Prediction, as often the relative stabilities of molecular crystals are approximated using static lattice energies rather than finite temperature free energy calculations\cite{Price-CSP,Day-CSP,Price-lattice-energy-CSP}. 



The most widely used techniques for the calculation of molecular crystals are empirical force-fields and density functional theory (DFT). These techniques have been very successfully applied and have significantly advanced understanding\cite{Beran_review_MolCrys,Tkatchenko_review_MolCrys,Kapil_MolCrys,Graeme-MolCrys,Price-force-field-for-trinitrobenzene,Johnson-XDM-for-X23,Price-CSP}, particularly when modern force-field parameterization and modern DFT exchange-correlation functionals are used. However, despite the success, the accuracy of these methods is not always clear and careful validation is required. Experiment and higher level electronic structure theories are the two obvious sources of validation. However, neither is entirely straightforward as direct like for like comparison with experiment is challenging (see below) and high level electronic structure references are scarce.
Indeed, so far each computation of a single lattice energy with a highly accurate correlated method represents a \textit{tour de force} study \cite{Sancho-Garcia-anthracene,Sherrill_CC_benzene,Chan-Science-Benzene, ZenPNAS2018,Klimes_RPAGSWSE,grergory_embCC,Chan-Science-Benzene,Hongo_DMC_Paradioodobenzene}, implying a lack of extensive high-accuracy reference values for molecular crystals and periodic solids in general.


Addressing this challenge, recent developments in electronic structure theory enabled accurate and efficient calculations for both surfaces and condensed phases\cite{Hutter_benzene_RPA,Beran_MP2_on_polymorphs,Berkelach_surface_chemistry,ZenPNAS2018}. Among these, diffusion Monte Carlo (DMC) is very promising for small and large molecules. DMC was shown\cite{ZenPNAS2018} to deliver lattice energies of molecular crystals at a computational cost comparable to the Random-Phase-Approximation (RPA) but with the accuracy of the so-called "gold standard" of quantum chemistry, coupled cluster with single, double, and perturbative triple excitations [CCSD(T)]. Specifically, DMC has been successfully applied to study 6 organic molecular crystals\cite{ZenPNAS2018,Hongo_DMC_Paradioodobenzene} as well as 13 ice polymorphs\cite{DMCICE13}, providing valuable insights into their energetics.

In this work, we consider the X23 dataset, the most used dataset for the lattice energies of molecular crystals comprising 23 materials. Very recent studies on X23 have shown that near chemical accuracy ($\sim 4 \text{ kJ/mol}$) can be achieved with second order Møller–Plesset perturbation theory (MP2) calculations \cite{Berkelbach_X23}, and that coupled-cluster methods achieve sub-chemical accuracy in the computation of the two-body terms\cite{Sherrill_X23_2body}. Here, we provide DMC reference computational values for the entire dataset. In addition, when comparison with previous state-of-the-art calculations is possible, we show that different high-accuracy computational methods agree on lattice energies within $\sim 4 \text{ kJ/mol}$, which is better than a sometimes larger disagreement among experiments. 
The feasibility and accuracy of DMC for large molecular crystals open up the road to lattice energies benchmarked directly against computed high-accuracy computational values as well as the production of reference values for more complex condensed phase systems.
%


\section{Results and discussion}\label{sec:Results}

\noindent We start by elucidating the difference between lattice energy and sublimation enthalpy, which is fundamental to the discussion presented throughout the manuscript. In assessing the relative stability of molecular crystals, simulations generally focus on computing the (zero temperature) lattice energy, defined as:
\begin{equation} 
E_\text{latt}= E_{\text{crys}}-E_{\text{gas}},
    \label{eq:lattice-energy}
\end{equation}
where $E_{\text{crys}}$ is the total energy per molecule in the crystal phase, and $E_{\text{gas}}$ is the total energy of the isolated molecule in the gas phase.
However, the physical quantity measured in experiments is the sublimation enthalpy. Experimental estimates of the lattice energy are then obtained by subtracting from measured sublimation enthalpies a computational vibrational term:
\begin{equation}
E^{\text{exp}}_\text{latt}= - \Delta H_\text{sub}^\text{exp}(T) + \Delta E^{\text{comp}}_\text{vib}(T),
\label{eq:experimental_lattice_energy}
\end{equation}
where  $E^{\text{exp}}_\text{latt}$ is the experimental lattice energy,  $\Delta H_\text{sub}^\text{exp}$ is the measured sublimation enthalpy at temperature $T$, and $\Delta E^{\text{comp}}_\text{vib}$ is the computational vibrational term, comprising both zero-point energy and thermal effects. It is important to mention here that $\Delta E^{\text{comp}}_\text{vib}$ is challenging to obtain from computation: the need for large periodic cells and the importance of anharmonicity in molecular crystals means that this is in general not affordable with reference ab initio methods\cite{Kapil_MolCrys}.

Overall, this means that reference lattice energies were so far extrapolated from experiments rather than computed with higher-level methods, which introduces deviations as a result of comparing an experiment at finite temperature to a simulated idealized model system.
Moreover, our analysis on the experimental sublimation enthalpy (see below) shows that deviations often larger than the chemical accuracy limit characterize the measured value of $\Delta H_\text{sub}^\text{exp}$. This introduces a (large) uncertainty on $E^{\text{exp}}_\text{latt}$, which is independent of the (additional) error on the vibrational computational contribution.

\begin{figure*}[tbh!]
    \centering
    \includegraphics[scale=0.5]{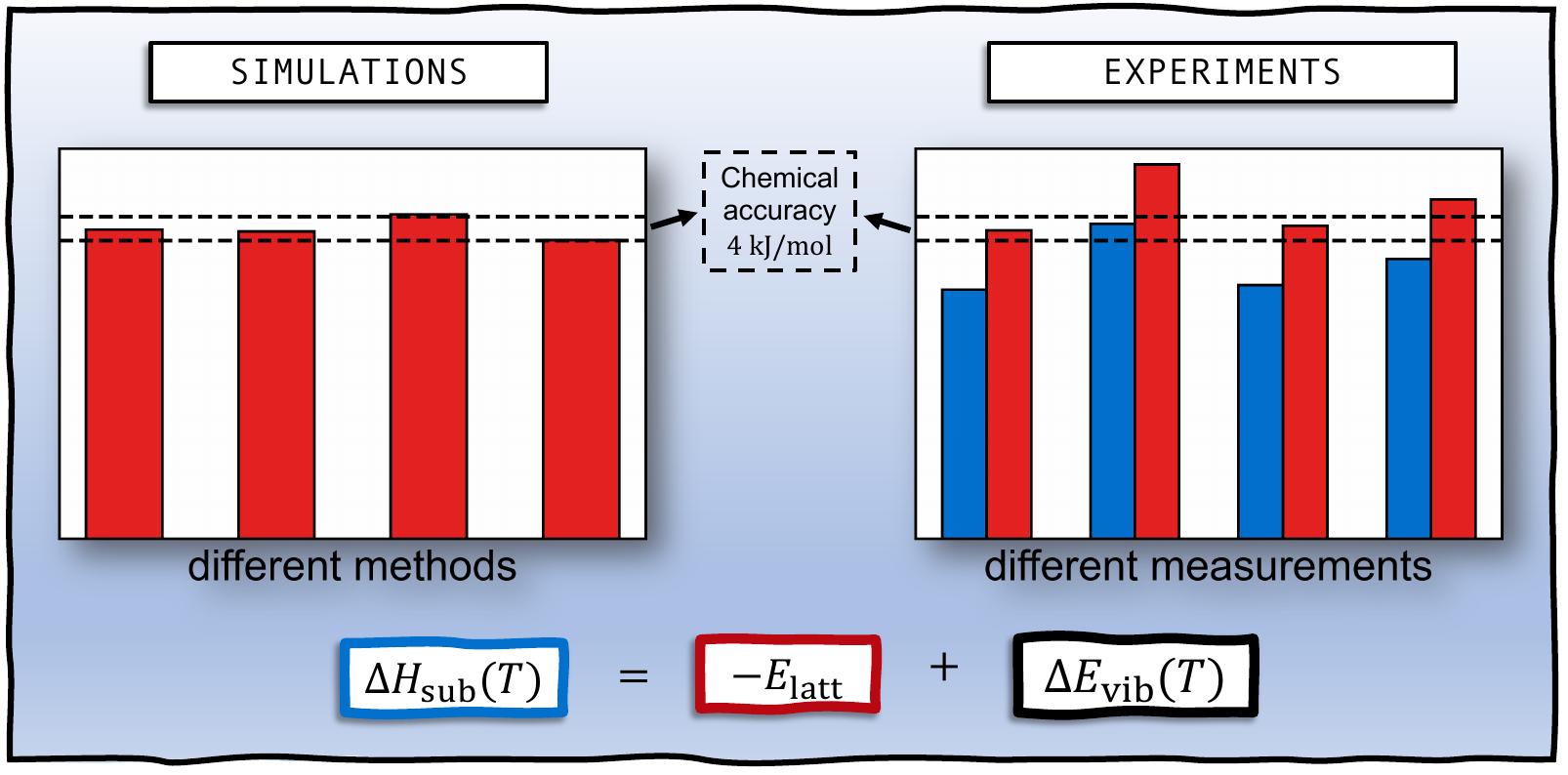}
    \caption{Schematic of the relation between the sublimation enthalpy $\Delta H_\text{sub}$ and the lattice energy $E_\text{latt}$. Simulations directly compute the lattice energy. Experimental estimates of the lattice energies $ E^{\text{exp}}_\text{latt}$ are obtained by subtracting a computational vibrational contribution $\Delta E^{\text{comp}}_\text{vib}$ from experimentally measured sublimation enthalpies $\Delta H^{\text{exp}}_\text{sub}$. Lattice energies and sublimation enthalpies are reported with red and blue bars, respectively. The left-hand-side illustrates that different high-accuracy computational methods agree on the estimate of the lattice energy within the chemical accuracy limit. This is opposed to the experimental scenario (right-hand-side) which can be characterized by larger uncertainties. The difference between the blue and red bars highlights that the lattice energy is the largest contribution ($\sim 80\%$) to the sublimation enthalpy.}
    \label{fig:cover}
\end{figure*}

In the following, as illustrated schematically in Fig.\ \ref{fig:cover}, we show that consensus within chemical accuracy is achieved on the lattice energy among explicitly correlated electronic structure methods, in stark contrast to a sometimes larger disagreement among experiments.

\subsection*{Consensus of computational methods on lattice energies}
\noindent
\begin{table}[tbh!]
\centering
\setlength{\tabcolsep}{1pt}
\renewcommand{\arraystretch}{1.3}
{
\begin{tabular}{lccl}
\hline
 & (DMC) & (DMC+DFT) & \\
Crystal               & Lattice energy  & Sublimation enthalpy &  \\ 
\hline \hline
1,4-cyclohexanedione & -88.3   $\pm$ 1.0           & 79.4 &  \\ 
Acetic acid          & -71.7    $\pm$ 0.6       & 65.7 &  \\
Adamantane           & -61.0     $\pm$  2.3          & 50.7 &  \\ 
Ammonia              & -38.2     $\pm$ 0.1         & 30.7 &  \\
Anthracene           & -100.2     $\pm$  0.5     & 91.7 &  \\ 
Benzene              & -49.8        $\pm$  0.2      & 39.9 &  \\ 
CO$_2$                  & -29.4     $\pm$  0.2         & 26.1 &  \\ 
Cyanamide            & -83.6        $\pm$  0.4        & 77.6 &  \\ 
Cytosine             & -156.2      $\pm$  1.0      & 149.0 &  \\ 
Ethyl carbamate      & -84.2      $\pm$  1.3      & 74.7 &  \\ 
Formamide            & -81.0       $\pm$  1.0        & 71.4 &  \\ 
Imidazole            & -88.2     $\pm$  0.8      & 79.3 &  \\ 
Naphthalene          & -75.5      $\pm$ 0.5         & 66.7 &  \\ 
Oxalic acid $\alpha$       & -102.6     $\pm$  1.4      & 97.6  &  \\
Oxalic acid $\beta$       & -102.3    $\pm$  0.6      & 99.0 &  \\ 
Pyrazine             & -61.1    $\pm$  1.1      & 53.2 &  \\
Pyrazole             & -77.3     $\pm$  0.5     & 70.9 &  \\ 
Triazine             & -60.5    $\pm$  0.6       & 53.6 &  \\ 
Trioxane             & -62.1    $\pm$  1.9       & 53.7 &  \\ 
Uracil               & -134.3    $\pm$  0.7      & 127.3 &  \\
Urea                 & -108.5        $\pm$  0.3      & 100.2 &  \\ 
Hexamine             & -86.2      $\pm$  0.6    & 76.9 &  \\
Succinic acid        & -125.2     $\pm$  0.5   & 118.2 &  \\
\cline{1-3}
\end{tabular}}
\caption{Lattice energies (kJ/mol) of the X23 molecular crystals computed with DMC. The reported error is the DMC statistical error bar. As discussed in Sec.\ \ref{sec:Results}, the overall error due to approximations involved in DMC and the DFT geometry optimization is estimated to be $\sim 2 \text{ kJ/mol}$. The second column reports the sublimation enthalpy (kJ/mol) at the temperature $T_\text{calc}$, estimated as the sum of the DMC lattice energy computed in this work and the DFT vibrational energies $\Delta E^{\text{comp}}_\text{vib}(T_\text{calc})$ computed in Ref.\ \cite{DHB_X23}. The temperature $T_\text{calc}$ is room temperature for every system except: acetic acid ($T_\text{calc}=290 \text{ K}$), ammonia ($T_\text{calc}=195 \text{ K}$), benzene ($T_\text{calc}=279 \text{ K}$), carbon dioxide ($T_\text{calc}= 207 \text{ K}$) and formamide ($T_\text{calc}=276 \text{ K}$).
As discussed in Sec.\ \ref{sec:Results}, the overall error on the sublimation enthalpy is estimated to be $\sim 6 \text{ kJ/mol}$.}
\label{tab:dmc_x23}
\end{table}

\noindent As described in equation \ref{eq:experimental_lattice_energy}, the experimental estimates of the lattice energy are extrapolated via a computational vibrational term.
For the X23 molecular crystals, the term $\Delta E^{\text{comp}}_\text{vib}$ has been previously computed with different approximations in Refs.\ \cite{OJ_C21,RT_X23,DHB_X23}. 
The most recent one, namely X23b\cite{DHB_X23}, was obtained using the quasi-harmonic approximation, averaging over four DFT functionals and taking into account both electronic and vibrational energy due to thermal expansion, and is therefore used in this work. 
In each of the previously reported datasets, a single initial value for the sublimation enthalpy was chosen to obtain reference experimental lattice energies. However, as discussed in Ref.\ \cite{DHB_X23}, the uncertainty on the initial value of the sublimation enthalpy can be larger than $\sim 5 \text{ kJ/mol}$.
To conduct a careful comparison, we consider all the values of the sublimation enthalpy $\Delta H_\text{sub}^\text{exp}$ reported in the literature (except those highlighted as unreliable\cite{Chickos_exp_sub_enth}), corrected with the X23b vibrational energy. 
The values of $\Delta H_\text{sub}^\text{exp}$ as a function of temperature were collected from Refs.\ \cite{Chickos_exp_sub_enth, succinic_acid_dasilva, hexamine_dekruif,nist_website} and are plotted in the Supporting Information (SI).
The vibrational terms $\Delta E^\text{comp}_\text{vib}$ were computed at a system specific temperature $T_\text{calc}$ (listed in Table 1 caption and in the SI). 
The values of $\Delta H ^\text{exp}_\text{sub}\left(T_\text{calc}\right)$ at the temperature $T_\text{calc}$ have been extrapolated according to the ideal approximation as described in the SI (section \ref{si:sec_exp_sub_enth}).

\begin{figure*}[tbh!]
    \centering
    \includegraphics[scale=.45]{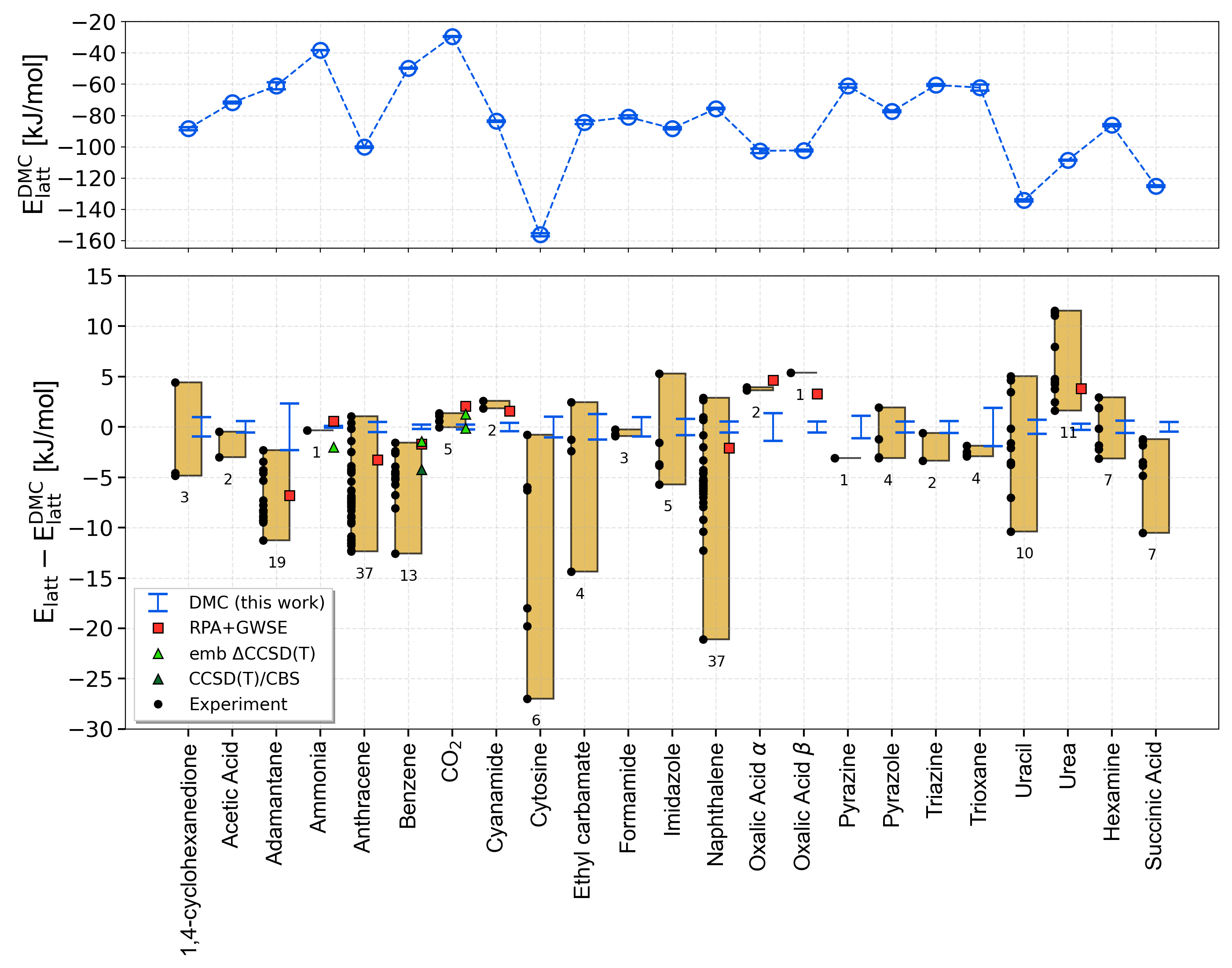}
    \caption{Performance of computations and experiments on the X23 lattice energies. (Top panel) DMC values of the electronic lattice energy for each system (the dashed line is to guide the eye). DMC has predictive accuracy in a large energy range going from $ -160 \text{ kJ/mol}$ to $ -20 \text{ kJ/mol}$. (Bottom panel) Difference between experimentally derived lattice energy (black dots) and DMC. The DMC statistical error bar is reported in blue. Experimental lattice energies are obtained by correcting experimental sublimation enthalpies with the most recent vibrational term (X23b), according to Eq.\ \ref{eq:experimental_lattice_energy}. The gold bar highlights the range of existing experimental measurements. The number of available experimental values is reported below each bar. Lattice energies obtained with RPA+GWSE (red squares) and CCSD(T) based methods (green triangles) are taken from Refs.\ \cite{Klimes_RPAGSWSE,grergory_embCC,Sherrill_CC_benzene}.}
    \label{figure1}
\end{figure*}

The X23 lattice energies computed with DMC are reported in Table \ref{tab:dmc_x23} and plotted in Fig.\ \ref{figure1}. In the upper panel, we plot the DMC lattice energies for each system, highlighting the variability of the X23 lattice energies over a relatively large energy range going from $-160 \text{ kJ/mol}$ to $-20 \text{ kJ/mol}$. The bottom panel shows for each system the difference between the experimental values (black dots) and the DMC lattice energies. The DMC statistical error bars are reported in blue. We also report values previously reported either with RPA with single corrections (RPA+GWSE) in Ref.\ \cite{Klimes_RPAGSWSE} (red squares) or with CCSD(T) methods (green triangles) in Refs.\ \cite{grergory_embCC,Sherrill_CC_benzene}, available only for a few systems.

The range in which the experimentally derived lattice energies vary is often larger than  $\sim 4 \text{ kJ/mol}$. In Fig.\ \ref{figure1}, this is evidenced by gold bars - under which the number of available measurements is reported - highlighting a current lack of consensus on the experimental value of the sublimation enthalpy for several molecular crystals. On the other hand, high-accuracy electronic structure methods generally agree within the chemical accuracy limit. Noticeable are the cases of anthracene, benzene, naphthalene, and urea, where computational methods agree within $\sim 4 \text{ kJ/mol}$, as opposed to the experimental uncertainties ranging from $\sim 10 $ to $\sim 25 \text{ kJ/mol}$. 

Overall, the `distance' between the experimental range (gold bar) and DMC is always within $\sim 4 \text{ kJ/mol}$ (with the only exception given by oxalic acid $\beta$, where only one experimental measurement is available), qualitatively validating the reliability of our estimates. Considering the spread in the experimental values of the lattice energy and the additional uncertainty on the necessary vibrational term (discussed below), we suggest that directly computed high-accuracy computational lattice energies have become at least as reliable as experimental ones, and could play a more significant role in benchmarking empirical and \textit{ab initio} methods.

\subsection*{Comparable uncertainties in experiments and simulations on sublimation enthalpies}
\noindent
So far we have focused on the performance of experiments and computation on the lattice energy. However, experimental estimates of the lattice energy involve the subtraction of a computational term. Therefore, we now address the accuracy of experiments and state-of-the-art simulations for sublimation enthalpies.

The experimental sublimation enthalpy, $\Delta H_\text{sub}^\text{exp}(T)$, is directly measured in experiments. Following the same procedure mentioned before and described in the SI, to allow for comparison with simulations, we extrapolated $\Delta H_\text{sub}^\text{exp}$ to the temperature $T_\text{calc}$ at which the vibrational contribution was available\cite{DHB_X23}.

We obtain the computational sublimation enthalpy, $\Delta H_\text{sub}^\text{comp}$, as \begin{equation}
    \Delta H_\text{sub}^\text{comp}\left(T_\text{calc}\right)= - E_\text{latt}^\text{DMC}+ \Delta E_\text{vib}^\text{DFT}\left( T_\text{calc} \right),
\end{equation}
where $ E^{\text{DMC}}_\text{latt}$ is the DMC lattice energy and $\Delta E^{DFT}_\text{vib}$ is the DFT vibrational thermal contribution computed in Ref.\ \cite{DHB_X23}. The values of  $\Delta H_\text{sub}^\text{comp}$ are reported in Table \ref{tab:dmc_x23}. 

The errors on experimental sublimation enthalpy are due to: (1) the spread in the reported measurement; and (2) the correction needed to extrapolate $\Delta H_\text{sub}^\text{exp}$ to the target temperature.
The range of variation of the computational sublimation enthalpy is due to: (1) errors in the computation of the lattice energy; and (2) errors in the computation of the vibrational term. The uncertainty on the DMC lattice energy is due to methods limitations (statistical error-bar, nodal surface, finite-size effect) and the considered geometry, optimized at the DFT level. We estimate the total error to be $\sim 2\text{ kJ/mol}$, as discussed in section \ref{si:sec_dmc_errors} of the SI. Sources of errors on the vibrational contribution are due to the inaccuracy of the DFT potential energy surface (PES) and the considered approximations (anharmonicity). Overall, an uncertainty of the order of $\sim 4 \text{ kJ/mol}$ on $\Delta E_\text{vib}^\text{DFT}$ has to be taken into account when comparing to experiments\cite{OJ_C21,RT_X23,DHB_X23}.
Finally, we get the total uncertainty by adding the system specific error on $\Delta E^{\text{DFT}}_\text{vib}$ reported in Ref.\ \cite{DHB_X23}.

\begin{figure*}[tbh!]
    \centering
    \includegraphics[scale=.55]{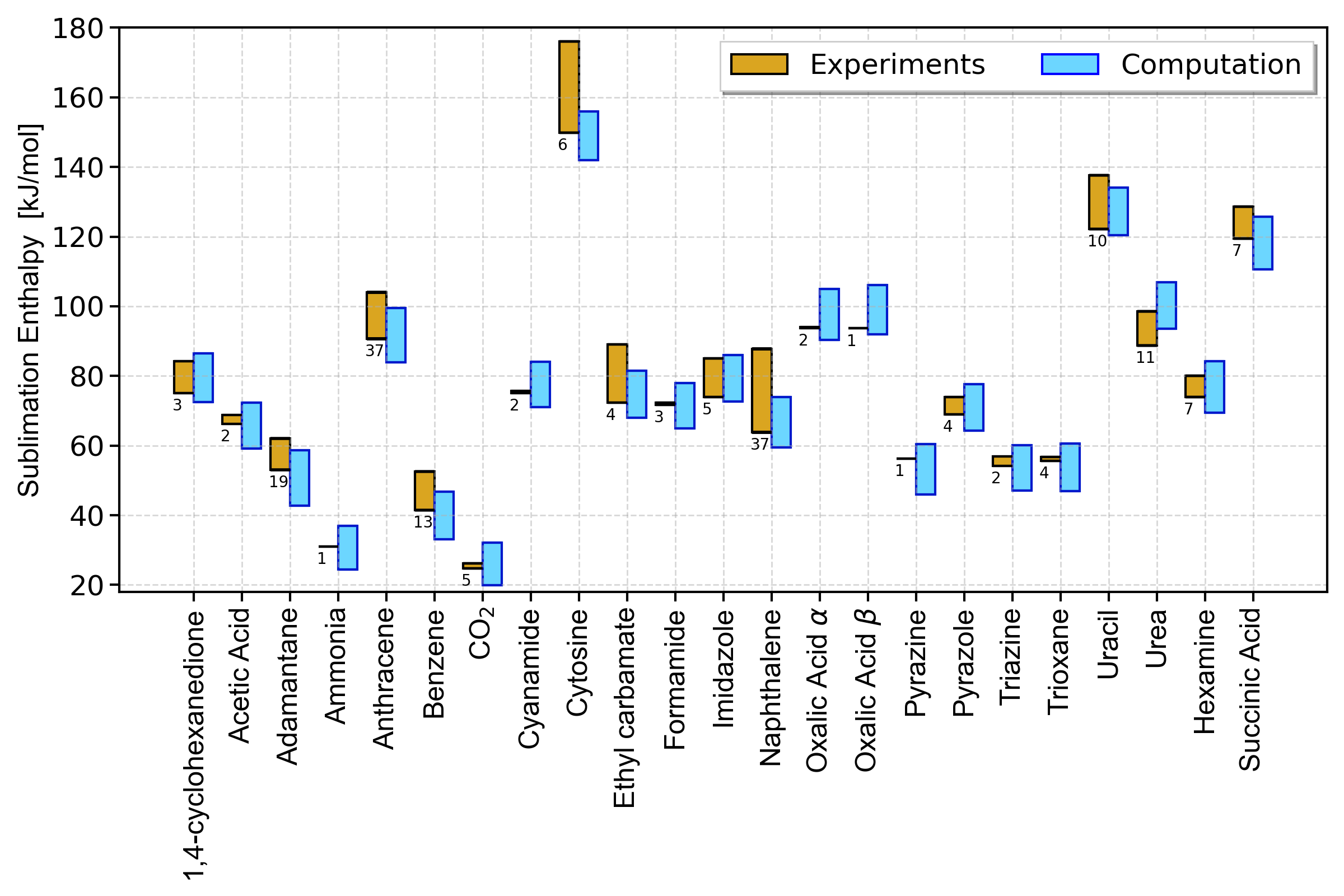}
    \caption{Comparison of uncertainties for experimental (gold) and  state-of-the-art computational (cyan) sublimation enthalpies. The gold bar is due to the spread in the literature sublimation enthalpies, extrapolated to the temperature $T_\text{calc}$ for which the computational vibrational contribution is available\cite{DHB_X23}. The number of available experimental values is reported below each bar. The cyan bar is due to geometry and methodological approximations used in the computation of electronic lattice energies and vibrational thermal contributions.}
    \label{figure2}
\end{figure*}

In Fig.\ \ref{figure2} we plot the estimated range for the X23 sublimation enthalpies in both experiments (gold) and computation (cyan).
Different from the lattice energy, it can be seen that state-of-the-art experimental and computational uncertainties for molecular crystals sublimation enthalpies are comparable. Moreover, they are overall larger than the sought after chemical accuracy limit. This poses an interesting question on whether the criterion usually considered to assess the quality of computational approaches is meaningful for current methods.
Overall, this work shows that to understand the accuracy of high-level computational methods on sublimation enthalpies, we would need both: (1) additional and accurate experimental measurements; and (2) to push the application of high-accuracy methods to the computation of fully anharmonic vibrational properties.

\section{Conclusions}\label{sec:Conclusions}
\noindent
This work has focused on lattice energies of molecular crystals. This quantity is not directly measured in experiments and corrections are needed for a direct comparison to simulations. On the other hand, high-accuracy explicitly correlated wave-functions methods were so far only applied to a few systems due to the demanding computational cost. Building on recent developments that enabled accurate and efficient diffusion Monte Carlo for large periodic unit cells, we computed the lattice energies of the X23 molecular crystals with DMC. The analysis of the performance of experiments and state-of-the-art simulations shows that, where direct comparison is possible, different high-accuracy computational methods have now reached consensus on the lattice energies within chemical accuracy. Larger uncertainties characterize instead experimental estimates. This work therefore provides valuable reference values for 23 molecular crystals. More generally, recent studies showed consensus of explicitly correlated methods for surface chemistry with chemical accuracy\cite{Benjamin_CO_MgO,Berkelach_CO_MgO,Berkelach_surface_chemistry,Gruneis_CCSDT_metallic}. Together with the advent of machine learning foundational models for material chemistry\cite{macemp0,nature_deepmind_foundational}, potentially requiring minimal data to fine-tune to a higher level of accuracy than the initial training, finite temperature simulations with the accuracy of explicitly correlated methods could soon become feasible.
This overall highlights an exciting time for future application of high-accuracy computational methods to both surfaces and condensed phases of more complex systems.
Finally, we have also shown that uncertainties of experimental and state-of-the-art computational sublimation enthalpies are currently comparable in magnitude. Our analysis suggests that the overall accuracy of sublimation enthalpy estimates could benefit from additional experiments as well as the application of higher-accuracy techniques to vibrational properties.

\section{Methods}\label{sec:Methods}
\noindent
Reference values for the lattice energies were computed with Fixed-Node DMC, using the CASINO\cite{CASINO} code. We use energy-consistent correlated electron pseudopotentials\cite{CASINO_PSEUDO_eCEPP} (eCEPP) with the most recent determinant locality approximation\cite{ZenDLA} (DLA). 
The trial wave functions were of the Slater-Jastrow type with single Slater determinants, and the single-particle orbitals obtained from DFT local-density approximation\cite{LDA} (LDA) plane-wave calculations performed with PWscf \cite{QuantumEspresso,PWscf} using an energy cut-off of $600 \text{ Ry}$ and re-expanded in terms of B-splines \cite{CASINO_BSPLINES}. The Jastrow factor included a two-body electron-electron (e-e) term, two-body electron-nucleus (e-n) terms, and three-body electron-electron-nucleus (e-e-n) terms. The variational parameters of the Jastrow have been optimized by minimizing the variance in the simulated cell for each analyzed crystal. The size of the simulation cell imposes some constraints on the Jastrow variational freedom, in the form of cut-offs in the e-n, e-e, and e-e-n terms. Following the workflow given in Ref.\ \cite{ZenPNAS2018}, tested on molecular crystals \cite{ZenPNAS2018} and ice polymorphs \cite{DMCICE13}, the simulation cells have been generally defined in order to guarantee the radius of the sphere inscribed in the Wigner-Seitz cell to be bigger than $5 \text{ Å}$.

The time step $\tau$ is a key factor affecting the accuracy of DMC calculations. In DMC, a propagation according to the imaginary time Schrödinger equation is performed to project out the exact ground state from a trial wave function \cite{QuantumMonteCarlo}. A time step $\tau$ must be chosen, but the projection is exact only in the continuous limit $\tau \to 0$. However, the ZSGMA \cite{ZSGMA} DMC algorithm gives better convergence with respect to $\tau$ than previously used methods because the time-step bias per molecule is independent of the size of the simulated cell in molecular crystals \cite{ZenPNAS2018}. In this work, we have verified the time step convergence for each analyzed system, as reported in the SI. We note that, in general, even in the limit of zero time step the DMC energy may be biased by the choice of the Jastrow factor, depending on how the non-local part of the pseudopotential is treated. This bias is eliminated if the DLA scheme is employed. 

The computation of $E_{\text{crys}}$ involves the use of periodic boundary conditions that can be subject to significant finite size errors (FSE). We took into account FSE using the Model Periodic Coulomb\cite{MPC1,MPC2,MPC3} (MPC) correction, and further correct for the (smaller) Independent Particle FSE (IPFSE) according to the procedure described in Ref.\ \cite{ZenPNAS2018}.

Further details on the DMC calculations, including the size of the supercell used in the solid phase calculations, the total energy of the solid phase with and without FSE corrections, and the total energy of the gas phase, are reported in the SI.

The geometries of the molecular crystals were optimized at the DFT level with the dispersion-inclusive functional optB88-vdW\cite{optPBE_optB88} using VASP\cite{VASP1,VASP2,VASP3,VASP4}, except for hexamine and succinic acid which were taken from Ref.\ \cite{RT_X23}. The k-point grid used for each system is reported in the SI. Tests on the effect of the geometry optimization on the DMC lattice energy are reported in section \ref{si:sec_dmc_errors} of the SI.

\section*{Supporting Information}
\noindent
See the Supporting Information for: (1) the experimental sublimation enthalpies reported in literature and extrapolated at the target temperature; (2) the analysis on the error on the vibrational thermal contribution; (3) details of the DMC calculations; (4) tests on the errors on the DMC lattice energies; (5) the geometries used in this study for both condensed and gas phases.

\section*{Acknowledgements}
\noindent
Calculations were  performed using the Cambridge Service for Data Driven Discovery (CSD3) operated by the University of Cambridge Research Computing Service (www.csd3.cam.ac.uk), provided by Dell EMC and Intel using Tier-2 funding from the Engineering and Physical Sciences Research Council (capital grant EP/T022159/1 and EP/P020259/1), and DiRAC funding from the Science and Technology Facilities Council (www.dirac.ac.uk). 
This work also used 
the ARCHER UK National Supercomputing Service (https://www.archer2.ac.uk), the United Kingdom Car Parrinello (UKCP) consortium (EP/ F036884/1). This research used resources of the Oak Ridge Leadership Computing Facility at the Oak Ridge National Laboratory, which is supported by the Office of Science of the U.S. Department of Energy under Contract No. DE-AC05-00OR22725).
AM
acknowledges support from the European Union under the “n-AQUA” European Research
Council project (Grant No. 101071937). D.A. and A.Z. acknowledges support from Leverhulme grant no. RPG-2020-038, and from the European Union under the Next generation EU (projects 20222FXZ33 and P2022MC742).



\section{Supporting Information for `How accurate are simulations and experiments for the lattice energies of molecular crystals?'}
\noindent
In the supporting information we provide:
\begin{itemize}
    \item the experimental sublimation enthalpies reported in literature and extrapolated to the target temperature (called $T_\text{calc}$ in the main manuscript) in section \ref{si:sec_exp_sub_enth};
    \item details of the DMC estimates of the lattice energy reported in the main manuscript in section \ref{si:sec_details-of-dmc};
    \item tests on the errors on the DMC lattice energies and on the time-step convergence in sections \ref{si:sec_dmc_errors} and \ref{si:sec_dmc_timestep};
    \item the k-point grid used in the geometry optimization in section \ref{si:k-point-grid},
    \item the geometry used in this study for each system in X23 in section \ref{si:geomeries}.

\end{itemize}

\subsection{Experimental sublimation enthalpies}\label{si:sec_exp_sub_enth}
\begin{figure*}[tbh!]
    \centering
    \includegraphics[scale=0.65]{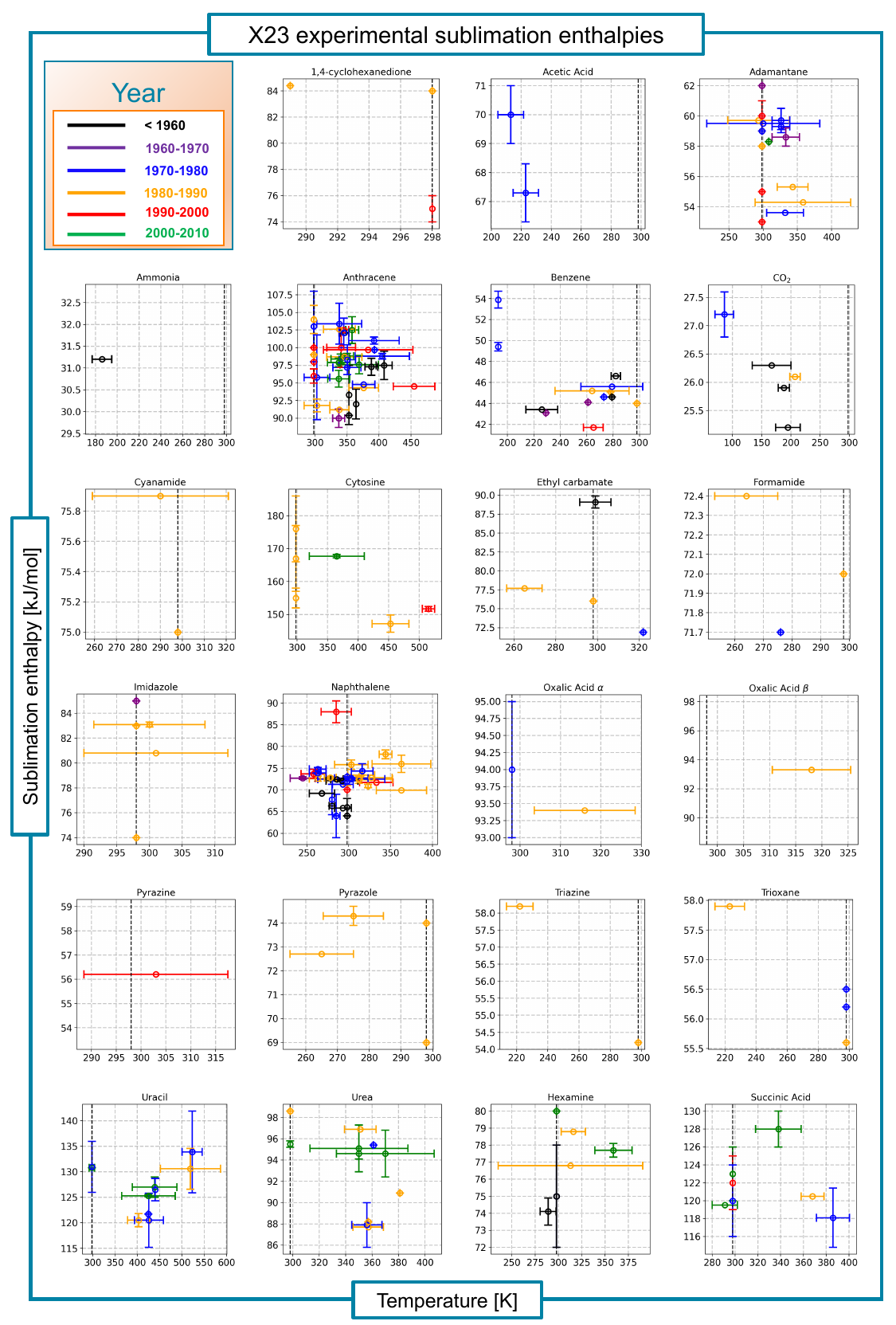}
    \caption{Experimental sublimation enthalpies reported in the literature as a function of the temperature. Points are color-coded according to the year of publication.}
        \label{fig:SI_exp_sub_enth}
\end{figure*}

In figure \ref{fig:SI_exp_sub_enth} we plot the literature values for the sublimation enthalpies of each molecular crystal as a function of the temperature. Data are collected from Refs.\ \cite{Chickos_exp_sub_enth, succinic_acid_dasilva, hexamine_dekruif, nist_website, OJ_C21, RT_X23, DHB_X23}. We explicitly report the plotted data in table \ref{tab:si-experimental-sublimation-enthalpies}. The points are color-coded according to the year of publication and the black dashed vertical line highlights room temperature. Figure \ref{fig:SI_exp_sub_enth} shows that the spread in the measured value is usually higher than $\sim 4 \text{ kJ/mol}$. We notice that cases where the experimental uncertainty shown in the main paper is small are usually due to a lack of data (see, for instance, oxalic acid, pyrazole, triazine).

\noindent The experimental lattice energies are obtained by subtracting the vibrational energy computed in Ref.\ \cite{DHB_X23}, according to the equation:
\begin{equation}
    E ^{\text{exp}}_\text{latt} = - \Delta H^\text{exp}_\text{sub}(T) + \Delta E^{\text{comp}}_\text{vib}(T).
\end{equation}
Since the vibrational energy is temperature dependent, we first need to extrapolate the experimental value of the sublimation enthalpy to the temperature $T_\text{calc}$ for which $ \Delta E^{\text{comp}}_\text{vib}$ was computed\cite{DHB_X23}.
The temperature $T_\text{calc}$ is room temperature for every system except: acetic acid ($T_\text{calc}=290 \text{ K}$), ammonia ($T_\text{calc}=195 \text{ K}$), benzene ($T_\text{calc}=279 \text{ K}$), carbon dioxide ($T_\text{calc}= 207 \text{ K}$) and formamide ($T_\text{calc}=276 \text{ K}$).

\begin{figure*}[tbh!]
    \centering
    \includegraphics[scale=0.6]{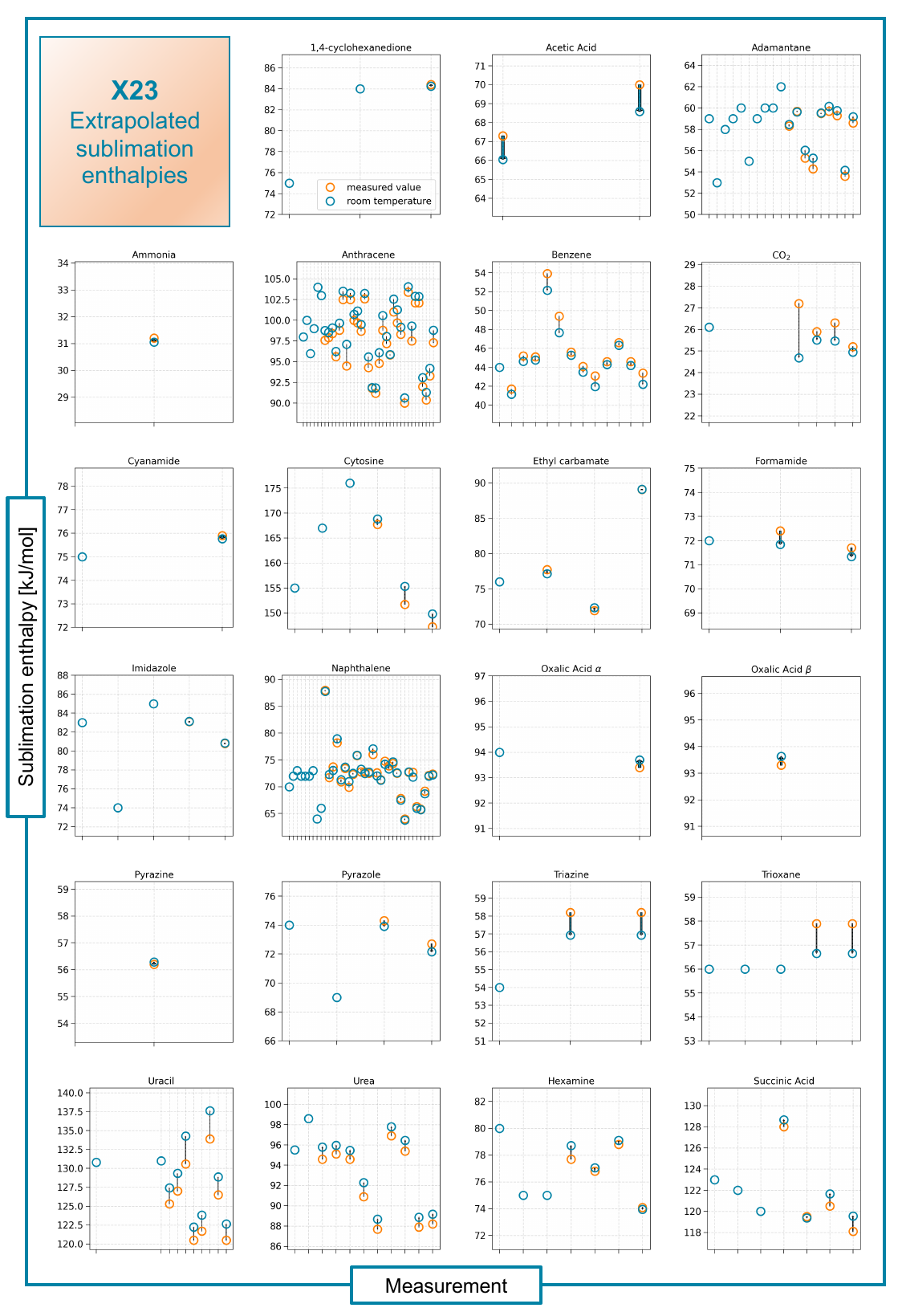}
        \caption{Experimental sublimation enthalpies extrapolated to the temperature $T_   \text{calc}$ according to equations 2 and 3. Orange circles are experimentally measured values at temperature $T\neq T_\text{calc}$. Cyan circles are the extrapolated values at temperature $T= T_\text{calc}$. Original and extrapolated values are connected by an arrow.}
            \label{fig:SI_exp_room_temp}
\end{figure*}

In figure \ref{fig:SI_exp_room_temp} we plot the values of the sublimation enthalpy extrapolated at the target temperature $T_\text{calc}$ using the ideal system approximation, i.e.,
\begin{align}
    \Delta H^{\text{exp}}_\text{sub} (T_\text{calc}) = \Delta H(T)- 2R(T_{\text{calc}}-T) \hspace{2cm} \text{non linear molecules,}\\
     \Delta H^{\text{exp}}_\text{sub} (T_\text{calc}) = \Delta H(T)- \frac{5}{2} R(T_{\text{calc}}-T) \hspace{2cm} \text{linear molecules.}
     \label{si:eq-extrapolation}
\end{align}
Orange circles represent values experimentally measured at a temperature $T$ different from the temperature $T_\text{calc}$ (i.e., the values reported in figure \ref{fig:SI_exp_sub_enth}); cyan circles are the extrapolated values at temperature $T_\text{calc}$. Original and extrapolated values are connected by an arrow.
We acknowledge that approaches more sophisticated than the ideal system correction are possible (i.e., evaluating the vibrational contribution in the quasi-harmonic approximation or use experimental heat capacity data). However, the spread in the measured values is such that the error coming from the ideal system approximation would be a minor correction to the analysis reported in the main paper.
 
Finally, in table \ref{tab:si-experimental-sublimation-enthalpies} we report the summary of the experimental sublimation enthalpy measurements considered in our analysis.
The acronyms of the experimental techniques are defined in Ref.\ \cite{Chickos_exp_sub_enth} and reported in table \ref{tab:si-acronysm-experiment} for clarity.

\begin{table}[tbh!]
\centering
\resizebox{\textwidth}{!}{
\begin{tabular}{ll} \hline \hline
ME       & Mass Effusion-Knudsen Effusion                                                    \\
TE       & Torsion Effusion                                                                  \\
A        & calculated from the vapor pressure data reported by the method of least   squares \\
TSGC     & Temperature Scanning Gas Chromatography                                           \\
BG       & Bourdon Gauge                                                                     \\
I        & Isoteniscope                                                                      \\
DBM      & Dibutyl Pththalate Manometer                                                      \\
CGC-DSC  & Combined Correlation Gas chromatography-Differential Scanning Calorimetry       \\
H        & Heat capacity                                                                     \\
MEM      & Modified Entrainment Method                                                       \\
GS       & Gas Saturation, transpiration                                                     \\
HSA      & Head Space Analysis                                                               \\
LE       & Langmuir Evaporation                                                              \\
RG       & Rodebush Gauge                                                                    \\
DSC      & Differential Scanning Calorimeter                                                 \\
MM       & mercury manometer                                                                 \\
GC       & Gas Chromatography                                                                \\
QR       & Quartz Resonator                                                                  \\
DM       & Diaphram manometer                                                                \\
KG       & Knudsen Gauge                                                                     \\
V        & Viscosity gauge                                                                   \\
E        & Estimated value                                                                   \\
QF       & Quartz Fiber                                                                      \\
MS       & Mass Spectrometry                                                                 \\
TPD      & Temperature Programmed Desorption                                                 \\
TPTD     & Temperature Programmed Thermal Expansion                                          \\ 
n/a       & not available                                                           \\ \hline \hline          
\end{tabular}}
\caption{Acronyms of the experimental methods defined in Ref.\ \cite{Chickos_exp_sub_enth} and used in table \ref{tab:si-experimental-sublimation-enthalpies}.}
\label{tab:si-acronysm-experiment}
\end{table}

\clearpage
\begin{longtable}{ccccc}
    \caption{Summary of experimental measurements of the sublimation enthalpy. For each system, the table reports respectively: the temperature at which the sublimation enthalpy is reported; the sublimation enthalpy (in kJ/mol, error reported in parentheses when available); the temperature range where the measurements was performed; the method used; the year of publication. The acronyms for the experimental methods are defined in Ref.\ \cite{Chickos_exp_sub_enth} and reported in table \ref{tab:si-acronysm-experiment}. When the temperature is not explicitly stated in Ref.\ \cite{Chickos_exp_sub_enth}, we assume for its value the average of the temperature range.} \\
    \label{tab:si-experimental-sublimation-enthalpies} 

\textbf{1,4-   Cyclohexanedione}               &          &                \\
Temperature       & Sublimation enthalpy & Temperature range & Method   & Year           \\ \hline \hline
289               & 84.4            & n/a               & C        & 1980-1990      \\
298               & 75(1)               & n/a               & TE       & 1990-2000      \\
298               & 84.2            & n/a               & ME       & 1980-1990      \\ \hline
\textbf{Acetic acid}       &                      &                   &          &                \\
Temperature       & Sublimation enthalpy & Temperature range & Method   & Year           \\ \hline \hline
223               & 67.3(1)             & 213-230           & TE,ME     & 1970-1980      \\
213               & 70(1)               & 213-230           & TE,ME     & 1970-1980      \\ \hline
\textbf{Adamantane}        &                      &                   &          &                \\ 
Temperature       & Sublimation enthalpy & Temperature range & Method   & Year           \\ \hline \hline
308               & 58.3            & n/a               & n/a      & 2000-2010      \\
293               & 59.7            & 278-368           & n/a      & 1980-1990      \\
343               & 55.3            & 328-373           & A        & 1980-1990      \\
358               & 54.3            & 343-483           & A        & 1980-1990      \\
300               & 59.5            & 278-443           & n/a      & 1970-1980      \\
326               & 59.7(0.8)            & 310-336           & TSGC     & 1970-1980      \\
326               & 59.3(0.2)            & 310-336           & BG       & 1970-1980      \\
332               & 53.6            & 312-366           & I        & 1970-1980      \\
333               & 58.6(0.6)            & 313-353           & DBM      & 1960-1970      \\
298               & 59.0            & n/a               & A        & 2000-2010      \\
298               & 53.0            & n/a               & CGC-DSC   & 1990-2000      \\
298               & 58.0            & n/a               & C        & 1980-1990      \\
298               & 59.6            & n/a               & H        & 1990-2000      \\
298               & 60.5(1)              & n/a               & H        & 1990-2000      \\
298               & 54.8            & n/a               & H        & 1990-2000      \\
298               & 59.3(0.2)            & n/a               & C        & 1970-1980      \\
298               & 59.5            & n/a               & n/a      & 1970-1980      \\
298               & 60.5(1)              & n/a               & H        & 1990-2000      \\
298               & 62.3            & n/a               & n/a      & 1960-1970      \\ \hline
\textbf{Ammonia}           &                      &                   &          &                \\
Temperature       & Sublimation enthalpy & Temperature range & Method   & Year           \\ \hline \hline
186               & 31.2            & 177-195           & n/a      & \textless 1960 \\  \hline
\textbf{Anthracene}        &                      &                   &          &                \\
Temperature       & Sublimation enthalpy & Temperature range & Method   & Year           \\ \hline \hline
369               & 97.6(1)              & 339-399           & ME       & 2000-2010      \\
337               & 97.9(0.6)            & 320-355           & n/a      & 2000-2010      \\
340               & 98.4(0.7)            & 320-350           & ME       & 2000-2010      \\
337               & 95.6(1)              & 320-354           & n/a      & 2000-2010      \\
350               & 98.8(0.4)            & 340-360           & ME       & 2000-2010      \\
358               & 102.5(2)             & 348-368           & ME       & 2000-2010      \\
455               & 94.5            & 423-488           & MEM      & 1990-2000      \\
345               & 102.5           & 338-353           & ME       & 1990-2000      \\
341               & 100.0(3)             & 318-363           & ME       & 1990-2000      \\
383               & 99.7            & 313-453           & GS       & 1990-2000      \\
346               & 98.7            & 318-373           & GS       & 1980-1990      \\
338               & 102.6           & 313-363           & GS       & 1980-1990      \\
376               & 94.3            & 353-399           & GS       & 1980-1990      \\
303               & 91.8(1)              & 283-323           & GS       & 1980-1990      \\
338               & 91.2            & 323-353           & GS       & 1980-1990      \\
376               & 94.8            & 358-393           & GS       & 1970-1980      \\
405               & 98.8(0.4)            & 363-448           & HSA      & 1970-1980      \\
350               & 97.2            & 328-372           & ME       & 1970-1980      \\
303               & 95.8(6)              & 283-323           & LE       & 1970-1980      \\
392               & 101.0(0.5)           & 353-432           & ME       & 1970-1980      \\
393               & 99.7            & n/a               & C        & 1970-1980      \\
350               & 98.3(2)              & 343-369           & n/a      & 1970-1980      \\
337               & 90.0(1)              & 327-346           & TE       & 1960-1970      \\
338               & 103.4(3)             & 303-373           & n/a      & 1970-1980      \\
408               & 97.5(2)              & 396-421           & HSA      & \textless 1960 \\
346               & 102.1           & 339-353           & n/a      & \textless 1960 \\
345               & 102.1(2)             & 338-353           & n/a      & 1970-1980      \\
364               & 92.0(2)              & n/a               & ME       & \textless 1960 \\
353               & 90.4            & n/a               & ME       & \textless 1960 \\
353               & 93.3(4)              & n/a               & n/a      & \textless 1960 \\
388               & 97.3(1)              & 378-398           & RG       & \textless 1960 \\
298               & 98.2            & 339-399           & ME       & 1990-2000      \\
298               & 100.2(0.4)           & n/a               & ME       & 1990-2000      \\
298               & 96.3(1)              & n/a               & DSC      & 1990-2000      \\
298               & 99.4            & n/a               & CGC-DSC   & 1980-1990      \\
298               & 104.5(2)             & n/a               & TE,ME     & 1980-1990      \\
298               & 102.9(5)             & n/a               & TE       & 1970-1980      \\ \hline
\textbf{Benzene}           &                      &                   &          &                \\
Temperature       & Sublimation enthalpy & Temperature range & Method   & Year           \\ \hline \hline
265               & 41.7            & 258-273           & n/a      & 1990-2000      \\
264               & 45.2            & 223-279           & A        & 1980-1990      \\
278               & 45.1            & n/a               & n/a      & 1980-1990      \\
193               & 53.9(0.8)            & n/a               & n/a      & 1970-1980      \\
193               & 49.4(0.4)            & n/a               & n/a      & 1970-1980      \\
279               & 45.6            & 221-268           & MM       & 1970-1980      \\
261               & 44.1            & n/a               & n/a      & 1960-1970      \\
229               & 43.1            & n/a               & n/a      & 1960-1970      \\
279               & 44.6            & n/a               & n/a      & \textless 1960 \\
282               & 46.6            & 263-270           & A        & \textless 1960 \\
273               & 44.6            & n/a               & n/a      & 1970-1980      \\
226               & 43.4            & 214-238           & A        & \textless 1960 \\
298               & 44.4            & 183-197           & TE,ME     & 1980-1990      \\ \hline
\textbf{CO$_2$}               &                      &                   &          &                \\
Temperature       & Sublimation enthalpy & Temperature range & Method   & Year           \\ \hline \hline
207               & 26.1            & 198-216           & A        & 1980-1990      \\
86                & 27.2(0.4)            & 70-102            & LE       & 1970-1980      \\
188               & 25.9            & 179-198           & n/a      & \textless 1960 \\
167               & 26.3            & 129-195           & A        & \textless 1960 \\
195               & 25.2            & 154-196           & n/a      & \textless 1960 \\ \hline
\textbf{Cyanamide}         &                      &                   &          &                \\
Temperature       & Sublimation enthalpy & Temperature range & Method   & Year           \\ \hline \hline
290               & 75.9            & 227-289           & TE,ME     & 1980-1990      \\
298               & 75.2            & n/a               & n/a      & 1980-1990      \\ \hline
\textbf{Cytosine}          &                      &                   &          &                \\
Temperature       & Sublimation enthalpy & Temperature range & Method   & Year           \\ \hline \hline
365               & 167.7(0.5)           & 320-410           & QR,ME     & 2000-2010      \\
515               & 151.7(0.7)           & 505-525           & GS       & 1990-2000      \\
453               & 147.2(3)             & 423-483           & ME       & 1980-1990      \\
298               & 155.0(3)             & n/a               & n/a      & 1980-1990      \\
298               & 167.0(10)            & n/a               & TE       & 1980-1990      \\
298               & 176.0(10)            & 450-470           & C        & 1980-1990      \\ \hline
\textbf{Ethyl carbamate}   &                      &                   &          &                \\
Temperature       & Sublimation enthalpy & Temperature range & Method   & Year           \\ \hline \hline
265               & 77.7            & 256-273           & TE,ME     & 1980-1990      \\
322               & 71.9            & n/a               & n/a      & 1970-1980      \\
299               & 89.1(0.8)            & 292-307           & GS       & \textless 1960 \\
298               & 76.0            & n/a               & n/a      & 1980-1990      \\ \hline
\textbf{Formamide}         &                      &                   &          &                \\
Temperature       & Sublimation enthalpy & Temperature range & Method   & Year           \\ \hline \hline
264               & 72.4            & 251-273           & TE,ME     & 1980-1990      \\
276               & 71.7            & n/a               & n/a      & 1970-1980      \\
298               & 71.7            & n/a               & n/a      & 1980-1990      \\ \hline
\textbf{Imidazole}        &                      &                   &          &                \\
Temperature       & Sublimation enthalpy & Temperature range & Method   & Year           \\ \hline \hline
300               & 83.1(0.2)            & 292-309           & ME       & 1980-1990      \\
301               & 80.8            & 288-310           & TE,ME     & 1980-1990      \\
298               & 83.1(0.2)            & n/a               & ME       & 1980-1990      \\
298               & 74.5(0.5)            & n/a               & C        & 1980-1990      \\
298               & 85.3            & n/a               & n/a      & 1960-1970      \\ \hline
\textbf{Naphthalene}       &                      &                   &          &                \\ 
Temperature       & Sublimation enthalpy & Temperature range & Method   & Year           \\ \hline \hline
285               & 88.0(2.5)            & 267-303           & ME       & 1990-2000      \\
333               & 71.7            & 313-353           & GS       & 1990-2000      \\
258               & 73.7(1)              & 243-273           & GS       & 1990-2000      \\
344               & 78.2(1)              & 337-352           & GC       & 1980-1990      \\
323               & 70.9(0.4)            & n/a               & DSC      & 1980-1990      \\
315               & 73.4            & 299-331           & GS       & 1980-1990      \\
363               & 69.9            & 333-393           & GS       & 1980-1990      \\
312               & 72.3(1)              & 293-331           & QR       & 1980-1990      \\
303               & 75.8(1)              & 283-323           & GS       & 1980-1990      \\
327               & 72.8            & 302-352           & GS       & 1980-1990      \\
278               & 72.8(0.3)            & 271-285           & ME       & 1980-1990      \\
313               & 72.5(0.1)            & 274-353           & DM       & 1980-1990      \\
363               & 76.0(2)              & 328-398           & DSC      & 1980-1990      \\
263               & 72.6(0.6)            & 253-273           & TE       & 1980-1990      \\
293               & 71.3            & 280-305           & GS       & 1970-1980      \\
263               & 74.8(0.4)            & 253-273           & TE       & 1970-1980      \\
263               & 73.9(0.2)            & 253-273           & ME       & 1970-1980      \\
316               & 74.3(2)              & 303-329           & TSGC     & 1970-1980      \\
303               & 72.5(0.3)            & 263-343           & DM       & 1970-1980      \\
280               & 67.8(3.5)            & n/a               & HSA      & 1970-1980      \\
285               & 64.0(5)              & 281-290           & LE       & 1970-1980      \\
303               & 72.7            & 283-323           & ME       & 1970-1980      \\
245               & 72.7(0.3)            & 230-260           & KG       & 1960-1970      \\
280               & 66.3            & 276-283           & V        & \textless 1960 \\
293               & 65.8            & 283-303           & Effusion & \textless 1960 \\
268               & 69.2            & 253-283           & n/a      & \textless 1960 \\
292               & 72.1            & 273-311           & n/a      & \textless 1960 \\
286               & 72.4            & 279-294           & n/a      & \textless 1960 \\
298               & 70.4            & n/a               & CGC-DSC   & 1990-2000      \\
298               & 72.3(0.4)            & n/a               & DSC      & 1980-1990      \\
298               & 72.6(0.1)            & n/a               & TE,ME,DM   & 1980-1990      \\
298               & 72.4(1)              & n/a               & C        & 1980-1990      \\
298               & 72.5            & n/a               & GS       & 1970-1980      \\
298               & 72.1(0.25)           & n/a               & C        & 1970-1980      \\
298               & 73.0(0.3)            & n/a               & C        & 1970-1980      \\
298               & 64.0            & n/a               & ME       & \textless 1960 \\
298               & 66.5(2)              & n/a               & QF       & \textless 1960 \\ \hline
\textbf{Oxalic acid alpha} &                      &                   &          &                \\
Temperature       & Sublimation enthalpy & Temperature range & Method   & Year           \\ \hline \hline
316               & 93.4            & 303-328           & n/a      & 1980-1990      \\
298               & 93.7(1)              & n/a               & TE       & 1970-1980      \\ \hline
\textbf{Oxalic acid beta}  &                      &                   &          &                \\
Temperature       & Sublimation enthalpy & Temperature range & Method   & Year           \\ \hline \hline
318               & 93.3            & 310-325           & n/a      & 1980-1990      \\ \hline
\textbf{Pyrazine}          &                      &                   &          &                \\
Temperature       & Sublimation enthalpy & Temperature range & Method   & Year           \\ \hline \hline
303               & 56.2            & 288-317           & n/a      & 1990-2000      \\ \hline
\textbf{Pyrazole}          &                      &                   &          &                \\
Temperature       & Sublimation enthalpy & Temperature range & Method   & Year           \\ \hline \hline
275               & 74.3(0.4)            & 268-287           & ME       & 1980-1990      \\
265               & 72.7            & 253-273           & TE,ME     & 1980-1990      \\
298               & 74.0(0.4)            & n/a               & n/a      & 1980-1990      \\
298               & 69.2(0.3)            & n/a               & C        & 1980-1990      \\\hline
\textbf{Triazine}          &                      &                   &          &                \\
Temperature       & Sublimation enthalpy & Temperature range & Method   & Year           \\ \hline \hline
222               & 58.2            & 212-229           & TE,ME     & 1980-1990      \\
298               & 54.2(0.2)            & n/a               & n/a      & 1980-1990      \\ \hline
\textbf{Trioxane}          &                      &                   &          &                \\
Temperature       & Sublimation enthalpy & Temperature range & Method   & Year           \\ \hline \hline
223               & 57.9            & 212-231           & TE,ME     & 1980-1990      \\
298               & 55.6            & n/a               & n/a      & 1980-1990      \\
298               & 56.5            & n/a               & C        & 1970-1980      \\
298               & 56.2(0.2)            & n/a               & C        & 1970-1980      \\ \hline
\textbf{Uracil}           &                      &                   &          &                \\
Temperature       & Sublimation enthalpy & Temperature range & Method   & Year           \\ \hline \hline
425               & 125.3(0.2)           & 315-435           & QR,ME     & 2000-2010      \\
405               & 130.8           & 399-411           & ME       & 2000-2010      \\
439               & 127.0(2)             & 394-494           & TE       & 2000-2010      \\
519               & 130.6(4)             & 452-587           & TE,ME     & 1980-1990      \\
403               & 120.5(1)             & 378-428           & QR       & 1980-1990      \\
425               & 121.7           & n/a               & MS       & 1970-1980      \\
523               & 133.9(8)             & 500-545           & HSA      & 1970-1980      \\
440               & 126.5(2)             & n/a               & C        & 1970-1980      \\
426               & 120.5(5)             & 393-458           & LE       & 1970-1980      \\
298               & 131.0(5)             & 452-587           & TE,GS     & 1970-1980      \\ \hline
\textbf{Urea}              &                      &                   &          &                \\
Temperature       & Sublimation enthalpy & Temperature range & Method   & Year           \\ \hline \hline
370               & 94.6(2)              & 329-403           & ME       & 2000-2010      \\
350               & 95.1(2)              & 329-403           & ME       & 2000-2010      \\
350               & 94.6(0.5)            & n/a               & C        & 2000-2010      \\
381               & 90.9            & n/a               & n/a      & 1980-1990      \\
357               & 87.7            & 345-368           & n/a      & 1980-1990      \\
351               & 96.9            & 338-362           & TE,ME     & 1980-1990      \\
361               & 95.4            & n/a               & n/a      & 1970-1980      \\
356               & 87.9(2)              & 345-368           & n/a      & 1970-1980      \\
357               & 88.2            & n/a               & n/a      & 1980-1990      \\
298               & 95.5(0.3)            & 358-402           & GS       & 2000-2010      \\
298               & 98.6            & n/a               & n/a      & 1980-1990      \\ \hline
\textbf{Hexamine}         &                      &                   &          &                \\ 
Temperature       & Sublimation enthalpy & Temperature range & Method   & Year           \\ \hline \hline
359               & 77.7            & 338-378           & GS       & 2000-2010      \\
313               & 76.8            & 298-453           & A        & 1980-1990      \\
316               & 78.8            & 302-328           & TE,ME     & 1980-1990      \\
289               & 74.1            & 281-298           & TE       & \textless 1960 \\
298               & 80.0            & n/a               & GS       & 2000-2010      \\
298               & 75.0(3)              & n/a               & V        & 1970-1980      \\
298               & 75.0(3)              & n/a               & n/a      & \textless 1960 \\ \hline
\textbf{Succinic acid}     &                      &                   &          &                \\
Temperature       & Sublimation enthalpy & Temperature range & Method   & Year           \\ \hline \hline
338               & 128.0(2)             & 318-358           & TPD      & 2000-2010      \\
291               & 119.5           & 280-302           & TPTD     & 2000-2010      \\
368               & 120.5           & 356-376           & TE,ME     & 1980-1990      \\
386               & 118.1(3)             & 372-401           & ME       & 1970-1980      \\
298               & 123.0(3)             & n/a               & n/a      & 2000-2010      \\
298               & 122.0(3)             & n/a               & n/a      & 1990-2000      \\
298               & 120.0(4)             & n/a               & n/a      & 1970-1980     
\end{longtable}

\clearpage

\subsection{Details of the DMC estimates of the lattice energy}\label{si:sec_details-of-dmc}
In table \ref{tab:si:details-dmc} we report further details on the estimates of the X23 lattice energies with DMC. In particular, we report for each system: (1) the size of the supercell used in the DMC calculation; (2) the number of electrons per molecule, $N_\text{e}$; (3) the number of molecules in the unit cell, $N_\text{mol}$; (4) the total energy of the solid with the Ewald summation; (5) the total energy of the solid with the Model Periodic Coulomb (MPC) correction; (5) the Independent Particle Finite Size Error (IPFSE); and (5) the total energy of the gas phase.
\begin{table}[tbh!]
\centering
\resizebox{\textwidth}{!}{
\begin{tabular}{cccccccc}
Crystal              & Cell  & $N_\text{e}$                         & $N_\text{mol}$ & Ewald (eV/unit cell) & MPC (eV/unit cell) & IPFSE (eV/unit cell) & Gas phase (eV/atom) \\ \hline \hline
1,4-cyclohexanedione & 2x2x2 & 44 & 2    & -3858.690 (0.019)    & -3858.184 (0.020)  & 0.0041               & -1928.179 (0.002)  \\
Acetic acid          & 1x3x2 & 24                         & 4    & -4986.300 (0.022)    & -4985.664 (0.023)  & 0.0013               & -1245.673 (0.002)  \\
Adamantane           & 2x2x2 & 56                         & 2    & -3616.719 (0.046)    & -3616.198 (0.048)  & 0.0197               & -1807.476 (0.002)  \\
Ammonia              & 2x2x2 & 8                          & 4    & -1279.108 (0.002)    & -1278.718 (0.002)  & 0.0057               & -319.285 (0.001)   \\
Anthracene           & 1x2x1 & 66                         & 2    & -4655.195 (0.010)    & -4652.944 (0.009)  & -0.0418              & -2325.454 (0.003)  \\
Benzene              & 2x2x2 & 30                         & 4    & -4100.258 (0.008)    & -4099.797 (0.008)  & 0.0002               & -1024.433 (0.001)  \\
CO$_2$                  & 2x2x2 & 16                         & 4    & -4105.625 (0.009)    & -4105.159 (0.009)  & 0.0003               & -1025.986 (0.001)  \\
Cyanamide            & 2x2x2 & 16                         & 8    & -5818.815 (0.037)    & -5818.347 (0.032)  & 0.0012               & -726.427 (0.001)   \\
Cytosine             & 1x1x3 & 42                         & 4    & -7789.765 (0.039)    & -7788.162 (0.041)  & 0.0315               & -1945.429 (0.002)  \\
Ethyl carbamate       & 2x2x2 & 36                         & 2    & -3441.354 (0.026)    & -3440.894 (0.026)  & 0.0120               & -1719.580 (0.002)  \\
Formamide            & 3x2x2 & 18                         & 4    & -3640.269 (0.038)    & -3639.963 (0.039)  & 0.0020               & -909.152 (0.001)   \\
Imidazole            & 2x2x2 & 26                         & 4    & -4277.264 (0.035)    & -4276.749 (0.033)  & 0.0035               & -1068.274 (0.002)  \\
Naphthalene          & 2x2x2 & 48                         & 2    & -3352.167 (0.010)    & -3351.621 (0.010)  & 0.0066               & -1675.031 (0.003)  \\
Oxalic acid alpha    & 2x2x2 & 34                         & 4    & -8339.690 (0.054)    & -8339.118 (0.056)  & 0.0004               & -2083.716 (0.002)  \\
Oxalic acid beta     & 2x2x3 & 34                         & 2    & -4169.899 (0.011)    & -4169.514 (0.011)  & 0.0273               & -2083.711 (0.002)  \\
Pyrazine             & 2x2x3 & 30                         & 2    & -2445.925 (0.023)    & -2445.602 (0.023)  & -0.0025              & -1222.167 (0.001)  \\
Pyrazole             & 2x1x2 & 26                         & 8    & -8549.553 (0.042)    & -8548.540 (0.043)  & 0.0004               & -1067.766 (0.002)  \\
Triazine             & 1x2x2 & 30                         & 6    & -7933.725 (0.072)    & -7932.856 (0.034)  & 0.0354               & -1321.522 (0.002)  \\
Trioxane             & 1x2x2 & 36                         & 6    & -11205.590 (0.111)   & -11204.542 (0.117) & 0.0165               & -1866.782 (0.002)  \\
Uracil               & 1x1x3 & 42                         & 4    & -8386.464 (0.028)    & -8384.954 (0.028)  & 0.0211               & -2094.852 (0.002)  \\
Urea                 & 2x2x2 & 24                         & 2    & -2395.398 (0.005)    & -2394.889 (0.006)  & 0.0383               & -1196.339 (0.001)  \\
Hexamine             & 2x2x2 & 56                         & 1    & -2203.616 (0.006)    & -2203.079 (0.006)  & -0.0071              & -2202.178 (0.003)  \\
Succnic acid         & 2x1x2 & 46                         & 2    & -4921.200 (0.008)    & -4920.115 (0.009)  & -0.0017              & -2458.760 (0.002) \\ \hline
\end{tabular}
}
    \caption{Details of the DMC calculations of the lattice energies. The table reports respectively: name of the system; cell used in the DMC simulations; number of electrons $N_\text{e}$ per molecule; number of molecules $N_\text{mol}$ in the unit cell; Ewald energy; MPC energy; IPFSE; energy in the gas phase. The energies of the solid phases are in eV/unit cell. The statistical DMC error bar on each energy is reported in parentheses. All the energies are computed with a time step of $0.01 \text{ au}$.}
    \label{tab:si:details-dmc}
\end{table}

\clearpage

\clearpage
\subsection{Error on the DMC lattice energies }\label{si:sec_dmc_errors}
\begin{figure*}[tbh!]
    \centering
    \includegraphics[scale=0.45]{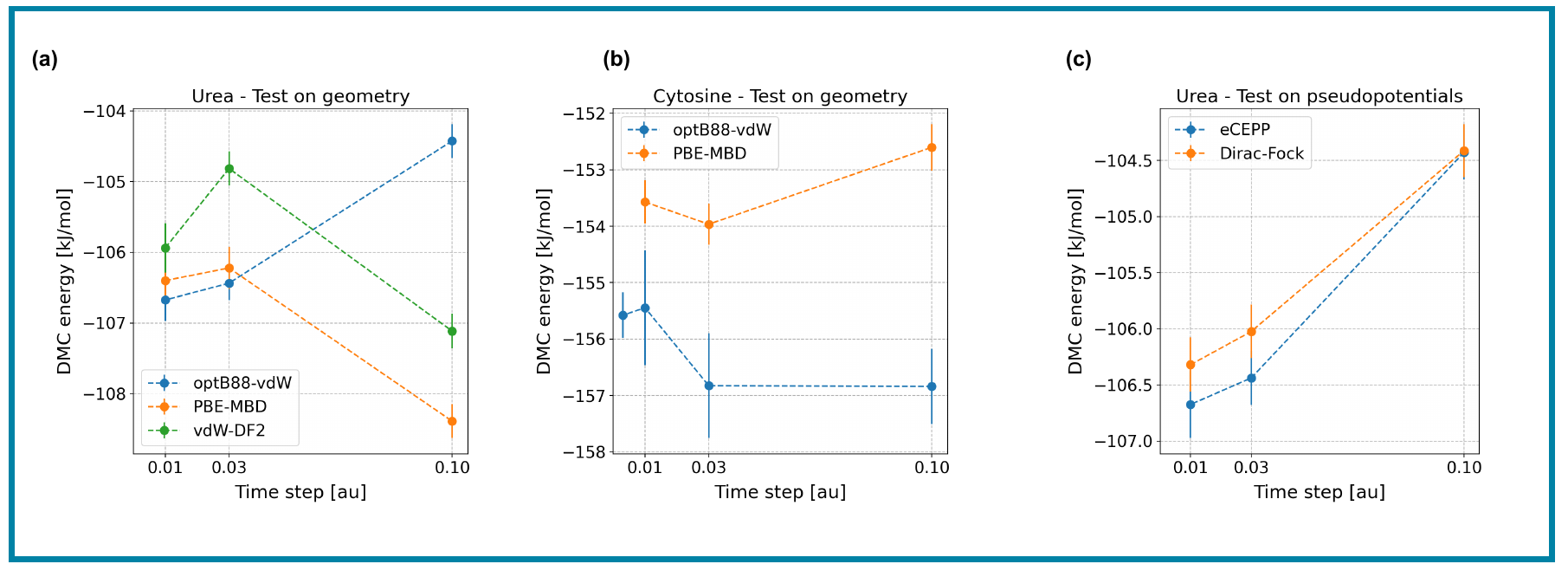}
    \caption{Panels (a,b): DMC lattice energy of urea (a) and cytosine (b) as a function of the time step for geometries optimized with different DFT functionals. (c) Test on the dependence of the DMC lattice energy of cytosine on the pseudopotentials: the results obtained with two different pseudopotentials, eCEPP and Dirac-Fock, agree within the DMC statistical error bar.
}
    \label{fig:SI-DMC-errors}
\end{figure*}

As discussed in the main paper, the main source of errors on the DMC estimates of the lattice energies are: (1) the considered geometry (because it is optimized at the DFT level); and (2) the pseudopotentials. We tested these errors on two showcase systems and estimate the total DMC error to be $\sim 1-2 \text{ kJ/mol}$. We report the error analysis in figure \ref{fig:SI-DMC-errors} as follows.
In panels \textbf{a} and \textbf{b} we compute the DMC lattice energy of urea and cytosine considering geometries optimized with different DFT functionals. We estimate the error due to the DFT geometry optimization to be $\sim 1-2 \text{ kJ/mol}$. In panel \textbf{c} we report the DMC lattice energy of urea with two different pseudopotentials, namely eCEPP\cite{CASINO_PSEUDO_eCEPP} (used in this work) and the Dirac-Fock \cite{CASINO_PSEUDO_1,CASINO_PSEUDO_2} used in a previous work \cite{ZenPNAS2018}. The DMC estimates (converged with respect to the time step) are consistent within the statistical error bar.

\clearpage
\subsection{Time step convergence of the DMC lattice energies}\label{si:sec_dmc_timestep}
\begin{figure*}[tbh!]
\centering
    \includegraphics[scale=.55]{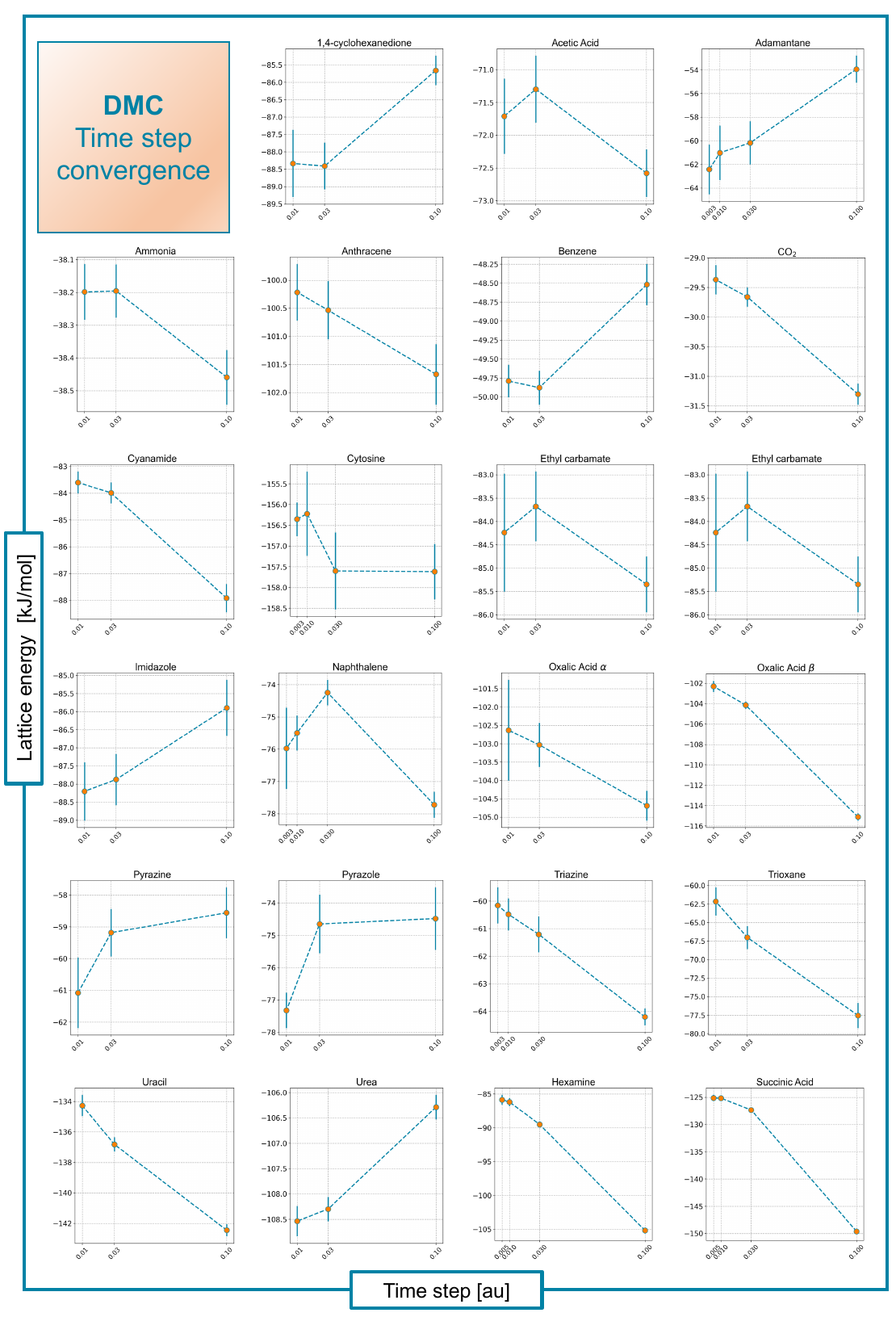}
    \caption{Convergence of the DMC estimates of the lattice energies with respect to the simulation time step. For each system, we plot the DMC lattice energy (in $\text{kJ/mol}$) as a function of the time step (in atomic units). The DMC values reported in the main manuscript are computed with a time step of $0.01 \text{ au}$.}
    \label{fig:SI-dmc-time-step}
\end{figure*}

The time step convergence is a key factor affecting the DMC estimates of the lattice energy, as discussed in the methodological section of the main paper. In figure \ref{fig:SI-dmc-time-step}, we report the DMC lattice energy as a function of the simulation time step for each molecular crystal. The DMC values reported in the main manuscript are computed with a time step of $0.01 \text{ au}$.

\clearpage
\subsection{DFT geometry optimization: k-point grid}\label{si:k-point-grid}
Table \ref{tab:si:k-point-grid} reports the k-point grid used in the geometry optimization of the C21 molecular crystals. The geometries of hexamine and succinic acid were instead taken from Ref.\ \cite{RT_X23}.

\begin{table}[tbh!]
    \centering
    \begin{tabular}{lc}
    Crystal & k-point \\
    \hline
    \hline
1,4-cyclohexanedione&4 $\times $ 4 $\times $ 4 \\
Acetic acid&2 $\times $ 6 $\times $4\\
Adamantane&4 $\times $4 $\times $3\\
Ammonia&5$\times $ 5$\times $ 5\\
Anthracene&3 $\times $4$\times $ 3\\
Benzene&3$\times $ 3 $\times $4\\
CO$_2$&4 $\times $4$\times $ 4\\
Cyanamide&4 $\times $4$\times $ 3\\
Cytosine&2 $\times $3 $\times $6\\
Ethyl carbamate&5$\times $ 4 $\times $4\\
Formamide&7 $\times $3$\times $ 4\\
Imidazole&3 $\times $5$\times $ 3\\
Naphthalene&3 $\times $4 $\times $4\\
Oxalic acid $\alpha$&4 $\times $3 $\times $4\\
Oxalic acid $\beta$&5 $\times $4$\times $ 5\\
Pyrazine&3 $\times $4 $\times $6\\
Pyrazole&3 $\times $2$\times $ 4\\
Trazine&3 $\times $3 $\times $4\\
Trioxane&3 $\times $3 $\times $3\\
Uracil&2 $\times $2$\times $ 8\\
Urea&4$\times $ 4 $\times $5\\
    \end{tabular}
    \caption{K-point grid used in the DFT geometry optimization of the C21 molecular crystals.}
    \label{tab:si:k-point-grid}
\end{table}

\clearpage
\subsection{Geometries input file}\label{si:geomeries}
In this section we report the geometries used in this study for the solid and gas phase of each system in X23. \newline
\begingroup
\obeylines
\noindent
\#1,4-cyclohexanedione solid
32
Lattice="6.65 0.0 0.0 0.0 6.21 0.0 -1.1717 0.0 6.76934"
C        2.33943089       1.04586116       3.02756362
C        3.13887313       4.15083595       3.74173847
C        3.14137928       1.05381535       1.74135694
C        2.33691542       4.15883532       5.02795636
C        2.29867793       1.33346282       0.48877110
C        3.17959951       4.43849035       6.28055129
C        1.08332498       0.43536293       0.40805506
C        4.39495512       3.54038448       6.36128494
C        0.37649894       0.15086096       1.71907703
C        5.10180566       3.25587766       5.05026518
C        0.83730031       1.05013608       2.87842371
C        4.64100946       4.15512845       3.89089757
H        3.96089164       1.78027943       1.83683828
H        1.51740513       4.88529816       4.93245801
H        3.60389202       0.05582253       1.67214589
H        1.87438580       3.16085041       5.09717249
H        1.90932203       2.36598386       0.54620435
H        3.56892295       5.47102280       6.22313664
H        1.71476999       1.23496380       6.33619097
H        3.76351179       4.33995724       0.43312965
H        0.57544710       5.30210115       1.95712011
H        4.90286638       2.19711347       4.81225456
H        5.94594506       0.24488577       1.55324000
H       -0.46764658       3.34991784       5.21610880
H        0.55431857       2.09483363       2.65179757
H        4.92397862       5.19982866       4.11753328
H        0.37110885       0.76904526       3.83221532
H        5.10722007       3.87403276       2.93710016
O        2.89744551       1.02711290       4.12245462
O        2.58085944       4.13202363       2.64683262
O       -0.47541052       6.15417747       6.12126120
O        5.95364385       3.04915853       0.64807731
\#1,4-cyclohexanedione molecule
16
Lattice="19.0 0.0 0.0 0.0 19.0 0.0 0.0 0.0 19.0"
C        8.88549500       8.18627000       9.31883000
C       10.38056500       8.38655000       9.03672000
C        9.95913500      10.72654000      10.02168000
C        8.78468500      10.74207000       9.03431000
C       10.95393500       9.60975000       9.73853000
C        8.04929500       9.41075000       8.97339000
H       10.97825500       7.51183000       9.31492000
H       10.53646500       8.53934000       7.95552000
H        8.73398500       7.99364000      10.39427000
H        9.58108500      10.56042000      11.04448000
H       10.49919500      11.67930000      10.04260000
H        9.15869500      10.94424000       8.01650000
H        8.06013500      11.53179000       9.26022000
H        8.47349500       7.32070000       8.78900000
O       12.12625500       9.68930000      10.05298000
O        6.87374500       9.33069000       8.67122000
\---------------------------------------------------
\#Acetic acid solid
32
Lattice="13.151 0.0 0.0 0.0 3.923 0.0 0.0 0.0 5.762"
C        4.40741914       2.74913621       1.21164337
C        8.74357321       1.17389375       4.09262644
C       10.98296382       3.13530616       1.21169907
C        2.16803416       0.78771224       4.09271522
C        5.39825986       2.42504225       2.28754074
C        7.75273569       1.49797023       5.16853109
C       11.97382209       3.45942900       2.28757823
C        1.17717269       0.46357601       5.16858919
H        4.19452344       3.72797287       5.30862135
H        8.95647340       0.19506169       2.42760164
H       10.77006915       2.15646549       5.30867882
H        2.38092565       1.76656395       2.42769936
H        4.91284832       1.87478714       3.09684732
H        8.23814574       2.04822169       0.21584139
H       11.48840010       0.08669600       3.09686908
H        1.66259297       3.83630400       0.21587789
H        6.20651838       1.81612928       1.85997857
H        6.94446882       2.10688253       4.74098133
H       12.78205678       0.14535254       1.85999429
H        0.36894322       3.77765588       4.74099344
H        5.85762740       3.34640838       2.66509913
H        7.29337796       0.57659672       5.54608280
H       12.43318067       2.53807833       2.66514894
H        0.71780489       1.38492438       5.54615861
O        4.90922284       3.49806769       0.23572895
O        8.24177516       0.42495788       3.11671120
O       11.48474579       2.38629810       0.23582311
O        1.66624824       1.53671580       3.11683815
O        3.23161621       2.36369990       1.19313692
O        9.91937265       1.55934591       4.07411440
O        9.80718018       3.52079449       1.19315582
O        3.34381995       0.40223756       4.07417882
\#Acetic acid molecule
8
Lattice="22.0 0.0 0.0 0.0 22.0 0.0 0.0 0.0 22.0"
C        9.70683000      11.13026000      11.00030000
C       11.20960000      11.11776000      11.00128000
O       11.93428000      12.08798000      11.00089000
O       11.69701000       9.83950000      11.00357000
H        9.35016000      12.16050000      11.00002000
H        9.32984000      10.60096000      10.11785000
H        9.32921000      10.60076000      11.88215000
H       12.67079000       9.91916000      11.00080000
\---------------------------------------------------
\#Adamantane solid
52
Lattice="6.639 0.0 0.0 0.0 6.639 0.0 0.0 0.0 8.918"
C        6.58155455       6.58155701       1.77900740
C        6.41734506       1.19108982       0.89108350
C        1.02232719       1.35546855       8.91359096
C        0.10676434       5.33302432       0.89108306
C        5.16864494       1.02233000       0.00440917
C        1.19108652       0.10676835       8.02691752
C        5.50178203       5.16864547       8.91359094
C        1.35546421       5.50178415       0.00440904
C        5.33302277       6.41734608       8.02691754
C        6.58155411       6.58155710       7.13899232
C        3.26205485       3.26205737       2.67999144
C        3.26205510       3.26205712       6.23800612
C        3.42626403       4.51058999       3.56791565
C        4.51058677       3.09784590       5.35008113
C        3.09784543       2.01352437       3.56791523
C        2.01352303       3.42626818       5.35008134
C        2.18228182       4.67496862       4.46340822
C        4.67496442       4.34182982       4.45458981
C        4.34182770       1.84914562       4.46340781
C        1.84914523       2.18228448       4.45458944
H        0.82123061       0.05744290       2.43499833
H        6.30172863       2.08544737       1.52708636
H        1.92070069       1.50152990       0.61686072
H        0.91664809       2.25503076       8.28349222
H        5.70287867       6.46667122       2.43499834
H        0.22238051       4.43866674       1.52708624
H        5.02258571       1.92069984       8.30113287
H        4.26908091       0.91665330       0.63450513
H        2.08544653       0.22238246       7.39091822
H        4.60340841       5.02258419       0.61686058
H        5.60746114       4.26908328       8.28349203
H        1.50152343       4.60341419       8.30113275
H        2.25502822       5.60746074       0.63450508
H        4.43866268       6.30173163       7.39091814
H        6.46666367       0.82123094       6.48299920
H        0.05744545       5.70288373       6.48299911
H        2.38337863       3.37694333       2.02400079
H        3.37694538       4.14073051       6.89399969
H        4.14073079       3.14717153       2.02400075
H        3.14716401       2.38338357       6.89399980
H        3.54188065       5.40494761       2.93191255
H        5.40494655       2.98223188       5.98608044
H        2.98222893       1.11916692       2.93191212
H        1.11916279       3.54188211       5.98608044
H        1.28390869       4.82102984       3.84213813
H        4.82102342       5.24020017       5.07586612
H        5.24020092       1.70308456       3.84213796
H        1.70308602       1.28391414       5.07586565
H        2.28796110       5.57453062       5.09350675
H        5.57452852       4.23615372       3.82449360
H        4.23614839       0.94958366       5.09350634
H        0.94958103       2.28796079       3.82449334
\#Adamantane molecule
26
Lattice="22.0 0.0 0.0 0.0 22.0 0.0 0.0 0.0 22.0"
C       11.00033000      11.00006000       9.22329000
C       10.99981000      10.99999000      12.77707000
C       11.16530000      12.24794000      10.10991000
C        9.75197000      11.16532000      11.89040000
C       10.83484000       9.75212000      10.10977000
C       12.24780000      10.83463000      11.89061000
C        9.92053000      12.41087000      11.00134000
C        9.58918000       9.92069000      10.99865000
C       12.07938000       9.58908000      11.00150000
C       12.41075000      12.07938000      10.99913000
H       10.12250000      11.11569000       8.56868000
H       10.88401000      10.12194000      13.43126000
H       11.87861000      10.88437000       8.56922000
H       11.11564000      11.87808000      13.43132000
H       11.28308000      13.13899000       9.47466000
H        8.86077000      11.28300000      12.52556000
H       10.71699000       8.86110000       9.47442000
H       13.13892000      10.71682000      12.52582000
H        9.02510000      12.55174000      10.37645000
H        9.44800000       9.02511000      11.62331000
H       12.97502000       9.44796000      10.37686000
H       12.55165000      12.97493000      11.62391000
H       10.02212000      13.31003000      11.62874000
H        8.69013000      10.02241000      10.37117000
H       11.97774000       8.68997000      11.62893000
H       13.30987000      11.97806000      10.37165000
\---------------------------------------------------
\#Ammonia solid
16
Lattice="5.1305 0.0 0.0 0.0 5.1305 0.0 0.0 0.0 5.1305"
N        1.04470891       1.04467975       1.04472578
N        3.60995847       1.52057361       4.08577516
N        4.08580811       3.60992798       1.52051951
N        1.52055891       4.08582692       3.60997900
H        1.86198440       1.39631721       0.53263233
H        1.39638414       0.53260446       1.86208330
H        0.53256637       1.86191453       1.39633557
H        4.42723342       1.16893701       4.59786767
H        3.96163521       2.03264968       3.26841562
H        3.09781668       0.70333959       3.73416376
H        3.73412397       3.09785652       0.70317001
H        4.59793363       4.42716029       1.16892118
H        2.03268616       3.26859071       3.96157834
H        0.70326905       3.73418703       3.09787790
H        1.16887653       4.59789554       4.42732765
H        3.26851588       3.96156632       2.03262046
\#Ammonia molecule
4
Lattice="20.0 0.0 0.0 0.0 20.0 0.0 0.0 0.0 20.0"
N        9.82043500       9.82050500       9.82051000
H       10.66326500      10.12949500       9.33673000
H       10.12951500       9.33672500      10.66327000
H        9.33673500      10.66327500      10.12951000
\---------------------------------------------------
\#Anthracene solid
48
Lattice="8.4144 0.0 0.0 0.0 5.9903 0.0 -6.4104 0.0 9.05607"
C       -1.64278629       0.15473549       3.34291581
C       -0.82774836       0.93576416       2.55997539
C       -0.40533504       0.49190276       1.27088277
C        0.41565004       1.27145567       0.44481548
C        7.58120584       5.17873552       0.81242098
C        6.73745618       4.39918051       1.65680431
C        6.33623125       4.86911648       2.88318636
C       -5.57720583       0.81156448       8.24364899
C        1.58834997       4.71884433       8.61125450
C        2.40933505       5.49839724       7.78518721
C       -4.73345617       1.59111950       7.39926567
C        2.83174837       5.05453585       6.49609459
C       -4.33223124       1.12118352       6.17288362
C        3.64678630       5.83556451       5.71315417
C       -0.56041402       3.14988618       5.71315393
C        2.56441402       2.84041385       3.34291606
C       -1.37545161       3.93091483       6.49609473
C        3.37945162       2.05938520       2.55997526
C       -1.79786502       3.48705332       7.78518757
C        3.80186503       2.50324672       1.27088242
C       -2.61884992       4.26660587       8.61125510
C        4.62284993       1.72369417       0.44481490
C       -1.37000589       2.18358588       8.24364882
C        3.37400590       3.80671415       0.81242118
C       -0.52625663       1.40403077       7.39926513
C        2.53025663       4.58626926       1.65680486
C       -0.12503183       1.87396707       6.17288300
C        2.12903183       4.11633295       2.88318700
H       -1.95525169       0.51408641       4.32354488
H       -0.49437099       1.91459330       2.90572735
H        0.74568187       2.25281061       0.79155651
H        6.41437812       3.41943849       1.30782470
H        5.68509061       4.25802909       3.51036065
H        1.25831814       3.73748940       8.26451346
H       -4.41037811       2.57086151       7.74824528
H        2.49837100       4.07570670       6.15034263
H       -3.68109059       1.73227091       5.54570933
H        3.95925171       5.47621359       4.73252511
H       -0.24794889       3.50923732       4.73252491
H        2.25194889       2.48106270       4.32354508
H       -1.70882924       4.90974402       6.15034314
H        3.71282925       1.08055601       2.90572686
H       -2.94888200       5.24796087       8.26451427
H        4.95288201       0.74233917       0.79155573
H       -0.20317849       0.42428874       7.74824482
H        2.20717849       5.56601129       1.30782518
H        0.52610877       1.26287955       5.54570871
H        1.47789123       4.72742047       3.51036129
\#Anthracene molecule
24
Lattice="25.0 0.0 0.0 0.0 22.0 0.0 0.0 0.0 20.0"
C        9.70814500       8.53389500       9.99697000
C       11.07889500       8.53406500      10.00191000
C       11.81211500       9.75713500      10.00117000
C       11.07957500      11.00333500       9.99515000
C        9.65442500      10.95759500       9.99026000
C        8.98762500       9.75985500       9.99104000
C       13.21132500       9.78972500      10.00587000
C       11.78851500      12.21014500       9.99437000
C       13.18784500      12.24273500       9.99904000
C       13.92026500      10.99656500      10.00492000
C       15.34546500      11.04238500      10.00967000
C       16.01246500      12.24008500      10.00877000
C       15.29176500      13.46595500      10.00303000
C       13.92096500      13.46576500       9.99820000
H       15.89636500      10.10276500      10.01423000
H       13.76386500       8.84969500      10.01026000
H        9.16195500       7.59306500       9.99738000
H       11.63208500       7.59581500      10.00630000
H        9.10357500      11.89719500       9.98577000
H        7.89983500       9.74040500       9.98735000
H       11.23587500      13.15021500       9.98997000
H       17.10016500      12.25958500      10.01261000
H       15.83786500      14.40693500      10.00258000
H       13.36776500      14.40398500       9.99404000
\---------------------------------------------------
\#Benzene solid
48
Lattice="7.39 0.0 0.0 0.0 9.42 0.0 0.0 0.0 6.81"
C        6.94634688       1.32481685       6.76645828
C        6.37040201       0.42621627       0.85894469
C        0.57262390       0.89722475       5.90790629
C        6.81737613       8.52277523       0.90209378
C        1.01959799       8.99378373       5.95105534
C        0.44365317       8.09518316       0.04354179
C        4.13865312       6.03481667       6.76645828
C        6.94634683       3.38518320       3.36145818
C        4.13865302       8.09518314       3.36145840
C        3.25134692       3.38518330       0.04354172
C        0.44365312       6.03481680       3.44854181
C        3.25134694       1.32481683       3.44854166
C        4.71459796       5.13621624       0.85894466
C        6.37040204       4.28378370       4.26394470
C        4.71459806       8.99378370       4.26394474
C        2.67540204       4.28378373       5.95105538
C        1.01959796       5.13621630       2.54605532
C        2.67540194       0.42621627       2.54605529
C        3.12237615       5.60722456       5.90790637
C        0.57262383       3.81277531       2.50290624
C        3.12237604       8.52277526       2.50290643
C        4.26762388       3.81277542       0.90209362
C        6.81737614       5.60722467       4.30709374
C        4.26762394       0.89722472       4.30709365
H        6.60387371       2.35960826       6.73054386
H        5.56968034       0.74691993       1.52593899
H        1.01434081       1.59339679       5.19625918
H        6.37565919       7.82660320       1.61374085
H        1.82031966       8.67308006       5.28406104
H        0.78612630       7.06039173       0.07945617
H        4.48112628       7.06960836       6.73054385
H        6.60387368       2.35039169       3.32554380
H        4.48112633       7.06039163       3.32554387
H        2.90887372       2.35039162       0.07945618
H        0.78612632       7.06960829       3.48445623
H        2.90887367       2.35960835       3.48445616
H        5.51531957       5.45691993       1.52593892
H        5.56968038       3.96308001       4.93093892
H        5.51531958       8.67308007       4.93093888
H        1.87468043       3.96308005       5.28406111
H        1.82031962       5.45691998       1.87906111
H        1.87468042       0.74691991       1.87906115
H        2.68065919       6.30339674       5.19625914
H        1.01434083       3.11660328       1.79125911
H        2.68065912       7.82660315       1.79125910
H        4.70934081       3.11660323       1.61374088
H        6.37565917       6.30339671       5.01874091
H        4.70934088       1.59339683       5.01874093
\#Benzene molecule
12
Lattice="22.0 0.0 0.0 0.0 22.0 0.0 0.0 0.0 22.0"
C       11.41535000       9.66832500      10.97953500
C       10.58467000      12.33172500      11.02047500
C       11.99903000      10.56740500      11.87250500
C       10.00100000      11.43261500      10.12750500
C       11.58370000      11.89910500      11.89295500
C       10.41635000      10.10087500      10.10701500
H       11.73936000       8.62955500      10.96361500
H       10.26066000      13.37044500      11.03640500
H       12.77822000      10.22990500      12.55304500
H        9.22178000      11.77011500       9.44696500
H       12.03887000      12.60043500      12.58950500
H        9.96113000       9.39956500       9.41049500
\---------------------------------------------------
\#CO2 solid
12
Lattice="5.624 0.0 0.0 0.0 5.624 0.0 0.0 0.0 5.624"
C        0.33737131       0.33737125       0.33737125
C        3.14937126       0.33737121       3.14937118
C        0.33737126       3.14937119       3.14937121
C        3.14937123       3.14937122       0.33737122
O        1.01211377       1.01211367       1.01211370
O        5.28662874       5.28662874       5.28662871
O        2.47462878       5.28662879       3.82411372
O        3.82411371       1.01211365       2.47462871
O        5.28662862       3.82411384       2.47462854
O        1.01211391       2.47462859       3.82411386
O        3.82411364       2.47462875       5.28662878
O        2.47462887       3.82411364       1.01211366
\#CO2 molecule
3
Lattice="20.0 0.0 0.0 0.0 20.0 0.0 0.0 0.0 20.0"
C       10.00061500       9.99922500      10.00163500
O       10.67966500       9.32001500       9.33519500
O        9.32033500      10.67998500      10.66480500
\---------------------------------------------------
\#Cyanamide solid
40
Lattice="6.856 0.0 0.0 0.0 6.628 0.0 0.0 0.0 9.147"
N        0.98678911       1.10121243       0.99077714
N        5.86921305       5.52682299       8.15622040
N        2.44123171       5.52683243       5.56427784
N        4.41477435       1.10120445       3.58272460
N        5.86923013       4.41522049       3.58273232
N        0.98677058       2.21281036       5.56426921
N        4.41477470       2.21281058       8.15622196
N        2.44121109       4.41522122       0.99079308
N        1.00418807       6.60938087       3.20381811
N        5.85180871       0.01865656       5.94318052
N        2.42381641       0.01859504       7.77727817
N        4.43220989       6.60940496       1.36974230
N        5.85183345       3.29537075       1.36974155
N        1.00416241       3.33266396       7.77725785
N        4.43217819       3.33259595       5.94314838
N        2.42384593       3.29539420       3.20386558
C        0.99084076       0.49271138       2.14750428
C        5.86515902       6.13532592       6.99949613
C        2.43716021       6.13530272       6.72102533
C        4.41885426       0.49271529       2.42598724
C        5.86515277       3.80670123       2.42594728
C        0.99084508       2.82133178       6.72105241
C        4.41881034       2.82132239       6.99944631
C        2.43717779       3.80669979       2.14757469
H        1.44155719       0.67055139       0.17058773
H        5.41443985       5.95747474       8.97641252
H        1.98648545       5.95746271       4.74414103
H        4.86951020       0.67058557       4.40287378
H        5.41447961       3.98459053       4.40291523
H        1.44151673       2.64344230       4.74408475
H        4.86953587       2.64349042       8.97641347
H        1.98645601       3.98454595       0.17058653
H        0.31571373       1.87180537       0.86440283
H        6.54028750       4.75622829       8.28259175
H        3.11223508       4.75627221       5.43788177
H        3.74375951       1.87175865       3.70910561
H        6.54028342       5.18579686       3.70912097
H        0.31571250       1.44223840       5.43787943
H        3.74367631       1.44220161       8.28258266
H        3.11232348       5.18582400       0.86441911
\#Cyanamide molecule
5
Lattice="22.0 0.0 0.0 0.0 22.0 0.0 0.0 0.0 22.0"
N       10.74996500      11.26105000      10.27863000
N       10.86924500      10.12084000      12.50888000
C       10.84030500      10.62328000      11.45798000
H       10.47311500      10.68276000       9.49112000
H       11.52688500      11.87916000      10.06258000
\---------------------------------------------------
\#Cytosine solid
5252
Lattice="13.044 0.0 0.0 0.0 9.496 0.0 0.0 0.0 3.814"
C       12.83107499       1.46376843       1.23794632
C        6.73492849       8.03221372       3.14494625
C        0.21291826       6.21178009       0.66907077
C        6.30910263       3.28421291       2.57606930
C        1.70547959       2.50687019       2.09939188
C        4.81650979       6.98914290       0.19237955
C       11.33851972       7.25483573       3.62161916
C        8.22751524       2.24114502       1.71462439
C        2.22843641       1.26156776       2.56715488
C        4.29357061       8.23443310       0.66015863
C       10.81557991       6.00955398       3.15384832
C        8.75044460       3.48643594       1.24685571
C        1.47708890       0.15122457       2.33346513
C        5.04491201       9.34477543       0.42646806
C       11.56690582       4.89921607       3.38753120
C        7.99910280       4.59677784       1.48054908
H       12.90497622       8.85152191       1.31903395
H        6.66103350       0.64446126       3.22603785
H        0.13902378       4.10353677       0.58797128
H        6.38298561       5.39246154       2.49497702
H        2.01521512       4.53678374       1.97546003
H        4.50677748       4.95923450       0.06845208
H       11.02877600       9.28474347       3.74554792
H        8.53725427       0.21125141       1.83855323
H        3.37048911       3.57895190       2.61472445
H        3.15149628       5.91705023       0.70771869
H        9.67349667       8.32692273       3.10628417
H        9.89254336       1.16907280       1.19928893
H        3.17784495       1.22154206       3.08849984
H        3.34415989       8.27446609       1.18150145
H        9.86617044       5.96952568       2.63249684
H        9.69985954       3.52647035       0.72550255
H        1.76095250       8.64035409       2.64050615
H        4.76106885       0.85564721       0.73350243
H       11.28307572       3.89233456       3.08049837
H        8.28294684       5.60365682       1.17350594
N        0.29137105       0.24386573       1.68094157
N        6.23064454       9.25213959       3.58792561
N       12.75265039       4.99184306       0.22608698
N        6.81336251       4.50414085       2.13308818
N        0.51536959       2.58763483       1.47237257
N        6.00661919       6.90836363       3.37936008
N       12.52862352       7.33562211       0.43464730
N        7.03740486       2.16036632       2.34165711
N        2.40229736       3.63827254       2.28936198
N        4.11967443       5.85773770       0.38234347
N       10.64166915       8.38624836       3.43164085
N        8.92435133       1.10974565       1.52466007
O       11.73477809       1.48662658       0.61793295
O        7.83123705       8.00935635       2.52494932
O        1.30922684       6.23463023       1.28907971
O        5.21277719       3.26134733       3.19606926
\#Cytosine molecule
13
Lattice="25.0 0.0 0.0 0.0 25.0 0.0 0.0 0.0 22.0"
C       11.24786000      12.78985500      11.64078500
C       13.17995000      11.81944500      10.75268500
C       13.71462000      13.07304500      10.30327500
C       12.95730000      14.17489500      10.55414500
H       11.18849000      14.85318500      11.39457500
H       13.52822000       9.82129500      10.92847500
H       14.83162000      10.70315500      10.18711500
H       14.66445000      13.14725500       9.78518500
H       13.25509000      15.17870500      10.26230500
N       11.77133000      14.04717500      11.19433500
N       12.02305000      11.69031500      11.38235500
N       13.87564000      10.67414500      10.50889500
O       10.16838000      12.78283500      12.21481500
\---------------------------------------------------
\#Ethyl carbamate solid
26
Lattice="5.051 0.0 0.0 1.61883 6.82155 0.0 -1.89881 -1.07775 7.2201"
C        4.94843622       5.03868645       2.64153873
C       -0.17728522       0.70496460       4.57865724
C        4.39308219       3.78967588       1.99505226
C        0.37805745       1.95398964       5.22510042
C        0.20252159       1.74281031       1.10683178
C        4.56847247       4.00089363       6.11328203
N        1.28873600       0.96878754       0.95087115
N        3.48237352       4.77479927       6.26928189
O        0.48145171       2.91027211       1.73336051
O        4.28965578       2.83335504       5.48677947
O        4.10717294       1.42486705       0.74292884
O        0.66391746       4.31882578       6.47703076
H        4.12460418       5.72648462       2.87648135
H        0.64658470       0.01720103       4.34375993
H        0.42570004       4.80065562       3.57368751
H        4.34549414       0.94290719       3.64645003
H        5.64428566       5.55594027       1.96830028
H       -0.87312095       0.18765022       5.25188585
H        3.69174724       3.26460847       2.65941690
H        1.07939921       2.47902940       4.56073336
H        3.88354726       4.00550209       1.04757892
H        0.88758288       1.73815583       6.17257783
H        2.22575902       1.33808247       1.11945121
H        2.54533410       4.40553251       6.10071505
H        1.18581231       0.11386032       0.39366181
H        3.58527219       5.62993968       6.82643819
\#Ethyl carbamate molecule
13
Lattice="22.0 0.0 0.0 0.0 22.0 0.0 0.0 0.0 22.0"
C       10.20352500      12.19071000      11.15459000
C       12.17506500      10.87041000      11.08975000
C       12.56559500       9.42006000      10.91172000
N        8.83649500      12.13133000      11.12503000
O       10.85947500      13.20349000      11.31353000
O       10.72612500      10.93578000      11.01325000
H        8.34400500      13.00833000      11.04008000
H        8.38117500      11.28726000      10.80669000
H       12.49306500      11.26833000      12.06130000
H       12.59878500      11.51269000      10.30793000
H       13.65599500       9.32217000      10.96549000
H       12.12410500       8.79651000      11.69710000
H       12.23315500       9.04133000       9.93870000
\---------------------------------------------------
\#Formamide solid
24
Lattice="3.604 0.0 0.0 0.0 9.041 0.0 -1.27456 0.0 6.87688"
C        0.86868335       0.49796107       1.76956771
C        1.46078838       8.54303893       5.10729956
C        0.29604336       5.01836482       1.66885272
C        2.03338466       4.02243212       5.20808014
O        1.39099161       8.40103458       1.70750974
O        0.93850318       0.63994271       5.16936693
O       -0.22627431       3.88050592       1.73093786
O        2.55569758       5.16032638       5.14603239
N        0.91200333       1.41044869       0.80489020
N        1.41747311       7.63054479       6.07200168
N        0.25269332       5.93089010       2.63353882
N        2.07671371       3.10994660       4.24339014
H        0.49812091       2.33599769       0.97933043
H        1.83132413       6.70497823       5.89760088
H        0.66662493       6.85646147       2.45909899
H        1.66278004       2.18437354       4.41784395
H        0.10070188       1.19157059       6.78576607
H        2.22876790       7.84943228       0.09111393
H       -0.21053509       5.71203241       3.52950647
H        2.53995202       3.32882087       3.34740403
H        0.33433629       0.81741912       2.68369519
H        1.99515332       8.22355938       4.19318480
H        0.83043791       5.33781192       0.75473835
H        1.49902295       3.70300371       6.12221516
\#Formamide molecule
6
Lattice="20.0 0.0 0.0 0.0 20.0 0.0 0.0 0.0 20.0"
C       10.58759000       9.98113000      10.01158000
O       11.34070000      10.70746000       9.38703000
N        9.23423000      10.12246000      10.05831000
H        8.65930000       9.49550000      10.60397000
H        8.79391000      10.88372000       9.55515000
H       10.94198000       9.11628000      10.61297000
\---------------------------------------------------
\#Imidazole solid
36
Lattice="7.582 0.0 0.0 0.0 5.371 0.0 -4.7433 0.0 8.56418"
C        0.33729660       1.16812487       1.53226658
C        2.50140339       4.20287511       7.03191345
C        4.87305375       3.85362491       2.74982372
C       -2.03435370       1.51737491       5.81435623
C        0.92261763       2.97476250       2.56311008
C        1.91608253       2.39623754       6.00106975
C        4.28773320       0.28926257       1.71897980
C       -1.44903310       5.08173723       6.84520007
C        1.63393529       2.94163032       1.38697445
C        1.20476490       2.42936951       7.17720539
C        3.57641534       0.25613076       2.89511573
C       -0.73771510       5.11486919       5.66906426
N        1.24794517       1.78824454       0.74211690
N        1.59075492       3.58275524       7.82206308
N        3.96240535       4.47374463       3.53997318
N       -1.12370514       0.89725534       5.02420685
N        0.11121190       1.85960718       2.64921081
N        2.72748814       3.51139296       5.91496913
N        5.09913854       4.54510687       1.63287919
N       -2.26043855       0.82589284       6.93130070
H       -3.10238840       1.43135509       8.39660305
H        5.94108843       3.93964462       0.16757695
H        3.56943875       4.11685543       4.44966700
H       -0.73073855       1.25414462       4.11451304
H       -0.13519440       0.22445223       1.28500474
H        2.97389433       5.14654777       7.27917538
H        5.34554468       2.90995214       2.99708540
H       -2.50684467       2.46104768       5.56709457
H        0.95135157       3.72625155       3.34275642
H        1.88734865       1.64474856       5.22142335
H        4.25899894       1.04075140       0.93933366
H       -1.42029886       4.33024836       7.62484620
H        2.36706420       3.62428203       0.97583871
H        0.47163606       1.74671768       7.58834109
H        2.84328669       0.93878230       3.30625128
H       -0.00458635       4.43221772       5.25792871
\#Imidazole molecule
9
Lattice="22.0 0.0 0.0 0.0 22.0 0.0 0.0 0.0 22.0"
C       10.34888500       9.82964500      11.00015500
C       12.05692500      11.13122500      10.99943500
C       10.96063500      11.95890500      10.99986500
N        9.87032500      11.11220500      11.00084500
N       11.66705500       9.80854500      10.99976500
H        8.89782500      11.39232500      10.99941500
H        9.69610500       8.96604500      10.99965500
H       13.10217500      11.41135500      10.99915500
H       10.85327500      13.03395500      10.99915500
\---------------------------------------------------
\#Naphthalene solid
36
Lattice="8.0846 0.0 0.0 0.0 5.9375 0.0 -4.91153 0.0 7.1003"
C       -0.94973915       0.11743516       2.34434104
C        4.12280915       5.82006484       4.75595897
C        0.08050915       3.08618512       4.75595941
C        3.09256057       2.85131481       2.34434067
C       -0.18069041       0.97674390       1.58828068
C        3.35376039       4.96075611       5.51201933
C       -0.68853999       3.94549402       5.51201981
C        3.86160968       1.99200591       1.58828026
C        0.20693460       0.63246278       0.26731594
C        2.96613539       5.30503723       6.83298406
C       -1.07616418       3.60121341       6.83298500
C        4.24923388       2.33628652       0.26731509
C       -3.92114814       1.49896467       6.56173323
C        7.09421811       4.43853534       0.53856676
C        3.05191766       4.46771482       0.53856792
C        0.12115207       1.46978513       6.56173219
C       -3.55163422       1.13139694       5.28458738
C        6.72470421       4.80610307       1.81571263
C        2.68240326       4.10014667       1.81571343
C        0.49066648       1.83735330       5.28458667
H       -1.24420859       0.40588482       3.35327114
H        4.41727857       5.53161518       3.74702886
H        0.37497846       3.37463495       3.74703023
H        2.79809125       2.56286501       3.35326987
H        0.13369636       1.94022491       1.98886185
H        3.03937358       3.99727512       5.11143815
H       -1.00292639       4.90897510       5.11143829
H        4.17599609       1.02852488       1.98886180
H       -3.61071009       2.45931442       6.97232726
H        6.78378002       3.47818560       0.12797274
H        2.74148016       5.42806396       0.12797299
H        0.43158959       0.50943599       6.97232709
H       -2.93800303       1.80454192       4.68314386
H        6.11107300       4.13295810       2.41715615
H        2.06877153       4.77329223       2.41715672
H        1.10429822       1.16420776       4.68314336
\#Naphthalene molecule
18
Lattice="25.0 0.0 0.0 0.0 25.0 0.0 0.0 0.0 20.0"
C       10.75943000      10.66175000       9.99986000
C       12.13611000      10.66289000       9.99967000
C       12.86236000      11.88117000       9.99974000
C       12.13744000      13.11907000      10.00005000
C       10.71974000      13.08188000      10.00038000
C       10.04504000      11.88190000      10.00018000
C       14.28022000      11.91822000       9.99954000
C       12.86373000      14.33727000      10.00022000
C       14.24050000      14.33842000      10.00010000
C       14.95482000      13.11832000       9.99971000
H       14.82925000      10.97742000       9.99944000
H       10.21474000       9.71991000       9.99991000
H       12.68835000       9.72395000       9.99945000
H       10.17105000      14.02285000      10.00056000
H        8.95724000      11.86783000      10.00049000
H       12.31181000      15.27634000      10.00053000
H       14.78552000      15.28009000      10.00015000
H       16.04276000      13.13255000       9.99967000
\---------------------------------------------------
\#Oxalic acid alpha solid
32
Lattice="6.548 0.0 0.0 0.0 7.844 0.0 0.0 0.0 6.086"
C        0.37583320       0.43232711       5.56714048
C        6.17216680       7.41167289       0.51885952
C        2.89816686       0.43232726       2.52414046
C        3.64983329       3.48967259       5.56714077
C        6.17216695       3.48967268       2.52414055
C        3.64983314       7.41167274       3.56185954
C        2.89816671       4.35432741       0.51885923
C        0.37583305       4.35432732       3.56185945
H        0.63014501       2.31842781       5.13526789
H        5.91785499       5.52557219       0.95073211
H        2.64385478       2.31842790       2.09226848
H        3.90414549       1.60357209       5.13526836
H        5.91785480       1.60357199       2.09226856
H        3.90414522       5.52557210       3.99373152
H        2.64385451       6.24042791       0.95073164
H        0.63014520       6.24042801       3.99373144
O        1.02885192       7.78797454       4.65332457
O        0.20000596       1.71027152       5.81226428
O        5.51914808       0.05602546       1.43267543
O        6.34799404       6.13372848       0.27373572
O        2.24514883       7.78797525       1.61032426
O        4.30285199       3.97802499       4.65332480
O        5.51914872       3.97802522       1.61032446
O        4.30285117       0.05602475       4.47567574
O        2.24514801       3.86597501       1.43267520
O        1.02885128       3.86597478       4.47567554
O        3.07399372       1.71027191       2.76926524
O        3.47400595       2.21172807       5.81226494
O        6.34799398       2.21172823       2.76926519
O        3.47400628       6.13372809       3.31673476
O        3.07399405       5.63227193       0.27373506
O        0.20000602       5.63227177       3.31673481
\#Oxalic acid alpha molecule
8
Lattice="20.0 0.0 0.0 0.0 20.0 0.0 0.0 0.0 20.0"
C       10.36587000       9.59569000       9.45311500
C        9.63408000      10.40374000      10.54673500
H        9.27416000       8.21411000      10.07709500
H       10.72505000      11.78589000       9.92217500
O       11.21197000      10.09969000       8.74899500
O        8.78803000       9.90032000      11.25100500
O        9.96469000       8.32590000       9.38345500
O       10.03446000      11.67365000      10.61542500
\---------------------------------------------------
\#Oxalic acid beta solid
16
Lattice="5.33 0.0 0.0 0.0 6.015 0.0 -2.36848 0.0 4.89289"
C       -1.59862081       0.01833967       4.88466362
C        4.56014081       5.99666033       0.00822638
C        3.37590051       3.02583970       2.45467197
C       -0.41438051       2.98916030       2.43821803
H        2.31293959       5.40473072       0.93895207
H        0.64858041       0.61026928       3.95393793
H        1.83282006       2.39723053       1.50749338
H        1.12869994       3.61776947       3.38539662
O       -0.98133805       0.63431893       4.01775147
O        1.29558040       5.36571883       0.99010253
O        3.94285805       5.38068107       0.87513853
O        1.66593960       0.64928117       3.90278747
O        2.75861824       3.64181890       3.32158390
O        0.20290176       2.37318110       1.57130611
O        2.85017950       2.35821859       1.45634322
O        0.11134050       3.65678142       3.43654678
\#Oxalic acid beta molecule
8
Lattice="20.0 0.0 0.0 0.0 20.0 0.0 0.0 0.0 20.0"
C       10.36587000       9.59569000       9.45311500
C        9.63408000      10.40374000      10.54673500
H        9.27416000       8.21411000      10.07709500
H       10.72505000      11.78589000       9.92217500
O       11.21197000      10.09969000       8.74899500
O        8.78803000       9.90032000      11.25100500
O        9.96469000       8.32590000       9.38345500
O       10.03446000      11.67365000      10.61542500
\---------------------------------------------------
\#Pyrazine solid
20
Lattice="9.325 0.0 0.0 0.0 5.85 0.0 0.0 0.0 3.733"
N        1.40499064       5.84998041       3.73297772
N        7.92001424       0.00001959       0.00002228
N        6.06753258       2.92509006       1.86644313
N        3.25751491       2.92492431       1.86655294
C        0.69836538       1.04337826       0.46254872
C        0.69832735       4.80661044       3.27041876
C        8.62667265       1.04335779       0.46257865
C        8.62663679       4.80664948       3.27047302
C        5.36087000       1.88165371       2.32903541
C        3.96423188       1.88162545       2.32909675
C        3.96416917       3.96835664       1.40396090
C        5.36081816       3.96839334       1.40389856
H        1.26652668       1.89307486       0.85085658
H        1.26645924       3.95691268       2.88207739
H        8.05852889       1.89309050       0.85094731
H        8.05845493       3.95690589       2.88216080
H        5.92917440       1.03196080       2.71733985
H        3.39616265       1.03191180       2.71748655
H        3.39586849       4.81805766       1.01566459
H        5.92888770       4.81809866       1.01552052
\#Pyrazine molecule
10
Lattice="22.0 0.0 0.0 0.0 22.0 0.0 0.0 0.0 22.0"
N       12.40806500      11.00001000      11.00021000
N        9.59193500      11.00003000      11.00019000
C       11.69754500       9.95150000      11.43245000
C       10.30240500       9.95149000      11.43252000
C       10.30245500      12.04847000      10.56756000
C       11.69764500      12.04854000      10.56762000
H       12.25709500       9.08625000      11.78912000
H        9.74341500       9.08610000      11.78931000
H        9.74341500      12.91390000      10.21069000
H       12.25707500      12.91372000      10.21078000
\---------------------------------------------------
\#Pyrazole solid
72
Lattice="8.19 0.0 0.0 0.0 12.588 0.0 0.0 0.0 6.773"
C        2.37238002       0.70669445       2.61975653
C        6.46740079      11.88136297       4.15317427
C        2.37238143       5.58728983       6.00626851
C        6.46741482       7.00067670       0.76670599
C        3.21047888       1.47171391       3.45257555
C        7.30547541      11.11630143       3.32038996
C        3.21047705       4.82226682       0.06609731
C        7.30549435       7.76570929       6.70689087
C        2.40361542       2.50055385       3.92012162
C        6.49859529      10.08747247       2.85283860
C        2.40359389       3.79344635       0.53364899
C        6.49862009       8.79453972       6.23933801
C        5.84191897       2.35717102       6.27863951
C        1.74697328      10.23080098       0.49432366
C        5.84197608       3.93683676       2.89216567
C        1.74695457       8.65119192       3.88083817
C        4.97070673       1.37375737       0.00735274
C        0.87575907      11.21423689       6.76562552
C        4.97073282       4.92023648       3.39386440
C        0.87573932       7.66777188       3.37913752
C        5.72638851       0.66428786       0.93058490
C        1.63142600      11.92369465       5.84237652
C        5.72638587       5.62971737       4.31711583
C        1.63140643       6.95829964       2.45589283
H        0.33388181       2.93712774       3.45397007
H        4.42886289       9.65094960       3.31899726
H        0.33386317       3.35688868       0.06747992
H        4.42888190       9.23108750       6.70550917
H        2.61802416       3.33652591       4.57458815
H        6.71300070       9.25149511       2.19839955
H        2.61800175       2.95747968       1.18811801
H        6.71301638       9.63051141       5.58487978
H        4.25276812       1.29728668       3.68354698
H        0.15778290      11.29069714       3.08943542
H        4.25275750       4.99668989       0.29705761
H        0.15778243       7.59129527       6.47591908
H        2.60116941      12.39469471       2.05639446
H        6.69619278       0.19337261       4.71653156
H        2.60118260       6.48727935       5.44289084
H        6.69620976       6.10068117       1.33007017
H        7.78136977       1.02587574       1.55639652
H        3.68640726      11.56210878       5.21654877
H        7.78138623       5.26818139       4.94293013
H        3.68639108       7.31987668       1.83007397
H        5.48106643      12.40663292       1.56174091
H        1.38611556       0.18136708       5.21124316
H        5.48104473       6.47537158       4.94825894
H        1.38608564       6.11263943       1.82475017
H        3.93715993       1.20228245       6.51128341
H        8.03221757      11.38573482       0.26171662
H        3.93718864       5.09168470       3.12478479
H        8.03219589       7.49628749       3.64822845
H        5.64637762       3.12299610       5.53698809
H        1.55141628       9.46499132       1.23597645
H        5.64646070       3.17100155       2.15051939
H        1.55141630       9.41701566       4.62248403
N        1.17467838       2.31769763       3.38131801
N        5.26966782      10.27037113       3.39162670
N        1.17465631       3.97630274       6.76785316
N        5.26968836       8.61166546       0.00514684
N        1.13306782       1.22732131       2.57208434
N        5.22807273      11.36077100       4.20083495
N        1.13307041       5.06668277       5.95860370
N        5.22809757       7.52128355       0.81438530
N        6.96009693       1.22528901       0.94517764
N        2.86512273      11.36265952       5.82776694
N        6.96010156       5.06872042       4.33170964
N        2.86511193       7.51932118       2.44129023
N        7.05384808       2.26635459       0.07941876
N        2.95888217      10.32161193       6.69353482
N        7.05387107       4.02767586       3.46594293
N        2.95887315       8.56037566       3.30705089
\#Pyrazole molecule
9
Lattice="22.0 0.0 0.0 0.0 22.0 0.0 0.0 0.0 22.0"
C       11.01591500      11.87821500      11.66172500
C       10.13161500      10.89700500      11.16415500
C       10.86960500      10.20272500      10.22384500
H       10.62191500       9.36725500       9.58343500
H        9.10345500      10.71911500      11.44618500
H       10.83242500      12.63274500      12.41656500
H       12.89654500      10.52869500       9.63866500
N       12.10053500      10.77938500      10.20966500
N       12.21825500      11.80898500      11.08057500
\---------------------------------------------------
\#Triazine solid
54
Lattice="9.647 0.0 0.0 -4.8235 8.35455 0.0 0.0 0.0 7.281"
C       -0.65235583       1.12991458       1.82024997
C        5.47585596       7.22463544       5.46074985
C        4.17114312       7.22463397       1.82024946
C        0.65235762       1.12991641       5.46075067
C        1.30471321       0.00000018       1.82025006
C        3.51878737       8.35454985       5.46074955
C        4.17093746       3.91467714       4.24717658
C        5.47567719       1.65480059       0.60667625
C        4.17132180       1.65479939       4.24717607
C        5.47606364       3.91467857       0.60667673
C        6.12824050       2.78507284       4.24717625
C        3.51875967       2.78507236       0.60667642
C       -0.65217675       6.69974979       6.67432374
C        0.65256302       4.43987253       3.03382330
C       -0.65256401       4.43987141       6.67432375
C        0.65217880       6.69975102       3.03382334
C        1.30474087       5.56947740       6.67432370
C       -1.30474020       5.56947714       3.03382359
H        8.44845324       2.07594425       1.82025026
H       -3.62495283       6.27860577       5.46074975
H        3.62495264       6.27860452       1.82024979
H        1.19854785       2.07594557       5.46075032
H       -2.42640502       8.35454989       1.82024993
H        7.24990554       0.00000011       5.46074993
H        3.62481769       4.86067365       4.24726822
H        6.02187490       0.70884918       0.60676776
H        3.62512506       0.70884813       4.24726774
H        6.02218264       4.86067459       0.60676831
H        7.22055738       2.78502755       4.24726785
H        2.42644271       2.78502761       0.60676828
H       -1.19837445       7.64570093       6.67423232
H        1.19868264       3.49387636       3.03373160
H       -1.19868258       3.49387533       6.67423194
H        1.19837551       7.64570193       3.03373186
H        2.39705770       5.56952212       6.67423186
H       -2.39705718       5.56952264       3.03373203
N        3.44794953       8.35454968       1.82025019
N        1.37555106       0.00000050       5.46074993
N       -4.13572481       7.16328837       1.82024958
N       -0.68777450       1.19126185       5.46075034
N        0.68777606       1.19126208       1.82025009
N        4.13572442       7.16328812       5.46074957
N        3.44795759       2.78455816       4.24723519
N        6.19904226       2.78455894       0.60673497
N        5.51152337       1.59374116       4.24723453
N        4.13598173       3.97625069       0.60673557
N        5.51101884       3.97625082       4.24723510
N        4.13547576       1.59374102       0.60673496
N       -1.37554199       5.56999124       6.67426528
N        1.37554286       5.56999203       3.03376449
N        0.68751824       4.37829896       6.67426461
N       -0.68802301       6.76080886       3.03376517
N        0.68802476       6.76080851       6.67426520
N       -0.68751835       4.37829937       3.03376475
\#Triazine molecule
9
Lattice="22.0 0.0 0.0 0.0 22.0 0.0 0.0 0.0 22.0"
C        9.91484000       9.91148500      10.99987000
C       11.78950000      11.14893500      10.99987000
C        9.78050000      12.15370500      10.99993000
H       12.87815000      11.21446500      10.99981000
H        9.42702000       8.93600500      10.99979000
H        9.17989000      13.06399500      10.99980000
N       11.25283000       9.92324500      10.99988000
N        9.12185000      10.98906500      10.99979000
N       11.11021000      12.30179500      11.00021000
\---------------------------------------------------
\#Trioxane solid
72
Lattice="9.32 0.0 0.0 -4.66 8.07136 0.0 0.0 0.0 8.196"
C        0.56242606       1.21971970       8.06617326
C        3.32304792       7.94752329       8.06617957
C       -3.88371403       6.97437796       8.06615026
C        3.88490521       6.97442470       3.96817326
C       -0.56123035       1.21975328       3.96817957
C       -3.32184309       7.94756738       3.96815026
C        5.22320752       3.90969515       2.60276186
C        3.32310867       2.56751274       2.60276308
C        5.43550948       1.59307845       2.60278213
C        3.88570980       1.59308021       6.70076186
C        4.09802396       3.90970595       6.70076308
C        5.99810888       2.56752960       6.70078213
C        0.56285368       6.60029444       5.33396527
C       -1.33692143       5.25764647       5.33397494
C        0.77573782       4.28371321       5.33393003
C       -0.77459352       4.28376700       1.23596527
C       -0.56171430       6.60034515       1.23597494
C        1.33806593       5.25769447       1.23593003
H        0.56889287       1.23152076       6.95995251
H        3.30960026       7.94727271       6.95996112
H       -3.87676292       6.96285105       6.95993453
H        3.87791860       6.96292374       2.86195251
H       -0.56773717       1.23152460       2.86196112
H       -3.30838493       7.94731100       2.86193453
H        5.22940201       3.92100379       1.49664673
H        3.31023327       2.56722229       1.49664747
H        5.44222006       1.58204116       1.49666753
H        3.87901348       1.58206130       5.59464673
H        4.09183781       3.92100160       5.59464747
H        6.01102275       2.56723670       5.59466753
H        0.56924717       6.61194364       4.22786142
H       -1.35019210       5.25745110       4.22786453
H        0.78257594       4.27232307       4.22782043
H       -0.78148528       4.27240547       0.12986142
H       -0.56818043       6.61193558       0.12986453
H        1.35134914       5.25746755       0.12982043
H        0.98609856       2.13494696       0.29091773
H        2.31860250       7.85680929       0.29091918
H       -3.30292666       6.14986448       0.29089202
H        3.30413172       6.14989978       4.38891773
H       -0.98489246       2.13498589       4.38891918
H       -2.31740006       7.85684731       4.38889202
H        5.64746374       4.82487170       3.02304796
H        2.31841988       2.47735361       3.02305364
H        6.01595670       0.76807791       3.02305733
H        3.30527208       0.76807513       7.12104796
H        3.67375964       4.82487187       7.12105364
H        7.00280363       2.47734763       7.12105733
H        0.98693826       7.51541780       5.75471944
H       -2.34145965       5.16733580       5.75470317
H        1.35621756       3.45890274       5.75466147
H       -1.35507099       3.45893715       1.65671944
H       -0.98577210       7.51545645       1.65670317
H        2.34261234       5.16738930       1.65666147
O        1.34879253       0.12379262       0.34052837
O        8.53896985       1.10511505       0.34060865
O        4.09400830       6.84133073       0.34055552
O        5.22718875       6.84137463       4.43852837
O        7.97242760       0.12383470       4.43860865
O        0.78224032       1.10517803       4.43855552
O        6.00914509       2.81310390       3.07319249
O        3.87982019       3.79647340       3.07320501
O        4.09284977       1.46074976       3.07319719
O        5.22835408       1.46073367       7.17119249
O        3.31206901       2.81309911       7.17120501
O        5.44137899       3.79647183       7.17119719
O        1.34895489       5.50387932       5.80402312
O       -0.78043047       6.48661686       5.80412394
O       -0.56684522       4.15112973       5.80404403
O        0.56798005       4.15119067       1.70602312
O       -1.34778797       5.50392446       1.70612394
O        0.78159503       6.48669769       1.70604403
\#Trioxane molecule
12
Lattice="22.0 0.0 0.0 0.0 22.0 0.0 0.0 0.0 22.0"
C        9.75943500      10.23301500      10.48981500
C       11.66973500      11.53522500      10.48909500
C        9.76111500      12.28259500      11.55955500
H       12.75628500      11.48690500      10.57994500
H        9.41112500      10.69645500       9.54435500
H        9.40757500       9.20382500      10.57990500
H        9.41107500      12.79617500      12.45662500
H        9.41144500      12.79398500      10.63977500
H       11.36588500      12.02818500       9.54337500
O       11.17737500      10.20522500      10.50813500
O       11.17926500      12.28295500      11.59037500
O        9.24371500      10.96230500      11.59172500
\---------------------------------------------------
\#Uracil solid
48
Lattice="11.938 0.0 0.0 0.0 12.376 0.0 -1.8771 0.0 3.1364"
C        1.78225622       2.59429758       0.00574313
C        8.27862796       9.78170004       3.13064003
C        2.30947595       8.78217817       3.13070298
C        7.75142237       3.59381922       0.00569702
C        2.08435392       0.14277921       0.11203804
C        7.97654185      12.23322079       3.02437106
C        2.00753325       6.33070352       3.02422188
C        8.05334757       6.04530388       0.11218645
C        3.49427918       0.34926312       0.25303856
C        6.56661499      12.02674350       2.88336387
C        0.59759831       6.53708848       2.88324882
C        9.46328674       5.83891832       0.25316402
C        3.97708648       1.62332073       0.25591104
C        6.08380215      10.75269009       2.88047582
C        0.11466880       7.81109519       2.88040657
C        9.94622034       4.56491532       0.25598942
H        5.03601501       1.86086943       0.34067352
H        5.02487175      10.51514907       2.79570684
H       -0.94426597       8.04859019       2.79567310
H       11.00515462       4.32742414       0.34071883
H        3.55800472       3.66468878       0.09183989
H        6.50287158       8.71132352       3.04454121
H        0.53370246       9.85246089       3.04448938
H        9.52719652       2.52354487       0.09191202
H       -1.59771100       1.18094491       3.05919882
H       11.65860468      11.19504198       0.07721175
H        5.68954995       7.36893421       0.07722164
H        4.37133923       5.00706054       3.05918654
H        4.14803206      11.86043705       0.33011604
H        5.91286997       0.51557388       2.80628931
H       -0.05613653       5.67221686       2.80614055
H       10.11701886       6.70379140       0.33026704
N        3.15416454       2.70636101       0.13227215
N        6.90671928       9.66964452       3.00410640
N        0.93755944       8.89415270       3.00405129
N        9.12333732       3.48185324       0.13234143
N        1.31181370       1.29183053       0.00666728
N        8.74907549      11.08416448       3.12973953
N        2.78001716       7.47975718       3.12972506
N        7.28087082       4.89623760       0.00668009
O       -0.84489623       3.56668612       3.03878249
O       10.90577388       8.80930236       0.09757908
O        4.93652294       9.75463598       0.09784350
O        5.12437551       2.62136045       3.03855840
O        1.51190013      11.39754441       0.07574834
O        8.54899792       0.97845444       3.06067651
O        2.58005529       5.20945840       3.06030010
O        7.48082512       7.16654161       0.07613823
\#Uracil molecule
12
Lattice="22.0 0.0 0.0 0.0 22.0 0.0 0.0 0.0 22.0"
C       10.18885000       9.77116500      11.00021000
C       12.29376000      11.15610500      11.00006000
C       11.45036000      12.33994500      10.99914000
C       10.10477000      12.21263500      10.99909000
H       12.11392000       9.08670500      10.99942000
H        8.48348000      10.89748500      10.99908000
H       11.93069000      13.31037500      10.99904000
H        9.43296000      13.06643500      10.99885000
N        9.49330000      10.98326500      10.99992000
N       11.56305000       9.94220500      11.00020000
O       13.51652000      11.13553500      11.00115000
O        9.62653000       8.68962500      11.00078000
\---------------------------------------------------
\#Urea solid
16
Lattice="5.565 0.0 0.0 0.0 5.565 0.0 0.0 0.0 4.684"
C        0.00000082       2.37694268       1.52725882
C        2.78249961       5.15943899       3.15674210
O        5.56499918       2.37694144       2.80034008
O        2.78249958       5.15943957       1.88366086
N        0.81701025       3.19393841       0.82848252
N        4.74799034       1.55994645       0.82848382
N        3.59951001       4.34244250       3.85551749
N        1.96548988       0.41143555       3.85551629
H        1.44737778       3.82432890       1.32525221
H        4.11762001       0.92955187       1.32525286
H        4.22987798       3.71205169       3.35874627
H        1.33512097       1.04182830       3.35874594
H        0.81111773       3.18805976       4.49317523
H        4.75388025       1.56582130       4.49317429
H        3.59361857       4.34832031       0.19082520
H        1.97137967       0.40556043       0.19082477
\#Urea molecule
8
Lattice="22.0 0.0 0.0 0.0 22.0 0.0 0.0 0.0 22.0"
C       11.13769000      11.49895000      11.00094000
O       11.74707000      12.55972000      11.00028000
N        9.78540000      11.40471000      10.69863000
N       11.73819000      10.28413000      11.30402000
H        9.24993000      10.65093000      11.11422000
H        9.31407000      12.30128000      10.68780000
H       11.35317000       9.44028000      10.89551000
H       12.75007000      10.32741000      11.31220000
\---------------------------------------------------
\#Hexamine solid
22
Lattice="-3.4763251 3.4763251 3.4763251 3.4763251 -3.4763251 3.4763251 3.4763251 3.4763251 -3.4763251"
H        2.63776255       0.95716255      -0.31903745
H       -0.31903745       2.63776255       0.95716255
H        4.95296255       0.95716255       0.95716255
H       -0.31903745       0.95716255       2.63776255
H        0.95716255       2.63776255      -0.31903745
H        0.95716255      -0.31903745       2.63776255
H        1.47666255       3.15726255       3.15726255
H        3.15726255       1.47666255       3.15726255
H        0.95716255       4.95296255       0.95716255
H        2.63776255      -0.31903745       0.95716255
H        0.95716255       0.95716255       4.95296255
H        3.15726255       3.15726255       1.47666255
C        2.00886255       0.31906255       0.31906255
C        0.31906255       2.00886255       0.31906255
C        5.58196255       0.31906255       0.31906255
C        0.31906255       0.31906255       2.00886255
C        0.31906255       5.58196255       0.31906255
C        0.31906255       0.31906255       5.58196255
N        1.18656255       1.18656255       1.18656255
N        2.92786255      -2.28973745       2.92786255
N        2.92786255       2.92786255      -2.28973745
N       -2.28973745       2.92786255       2.92786255
\#Hexamine molecule
22
Lattice="25.0 0.0 0.0 0.0 25.0 0.0 0.0 0.0 25.0"
H       14.29000000      11.54825000      11.29875000
H       12.82950000      11.03195000      10.40445000
H       10.86810000      10.33855000      11.60575000
H       10.86810000      10.33855000      13.39415000
H       14.29000000      11.54825000      13.70115000
H       12.82950000      11.03195000      14.59545000
H       10.86810000      12.41885000      10.40445000
H       10.86810000      13.96775000      11.29875000
H       14.29000000      13.62885000      12.49995000
H       12.82950000      14.66145000      12.49995000
H       10.86810000      12.41885000      14.59555000
H       10.86810000      13.96775000      13.70115000
C       13.18860000      11.55265000      11.30635000
C       11.23960000      10.86365000      12.49995000
C       13.18860000      11.55265000      13.69355000
C       11.23970000      12.93085000      11.30625000
C       13.18860000      13.62005000      12.49995000
C       11.23970000      12.93085000      13.69375000
N       12.71540000      10.82355000      12.49995000
N       12.71550000      12.95085000      11.27165000
N       10.71000000      12.24185000      12.49995000
N       12.71550000      12.95085000      13.72815000
\---------------------------------------------------
\#Succinic acid solid
112
Lattice="10.9317762 0.0 -0.0129284 0.0 8.7402311 0.0 -0.292297 0.0 10.2113924"
C        0.30811980       8.14946555       4.94731594
C        5.01171980       0.59066555       0.15191594
C        0.38111980       4.96076555       2.39451594
C        4.93861980       3.77936555       2.70481594
C        1.32351980       8.44676555       3.87141594
C        3.99631980       0.29346555       1.22781594
C        1.39651980       4.66356555       1.31851594
C        3.92321980       4.07666555       3.78071594
C        0.16197130       8.14946555      10.05301214
C        4.86557130       0.59066555       5.25761214
C        0.23497130       4.96076555       7.50021214
C        4.79247130       3.77936555       7.81051214
C        1.17737130       8.44676555       8.97711214
C        3.85017130       0.29346555       6.33351214
C        1.25037130       4.66356555       6.42421214
C        3.77707130       4.07666555       8.88641214
C        5.77400790       8.14946555       4.94085174
C       10.47760790       0.59066555       0.14545174
C        5.84700790       4.96076555       2.38805174
C       10.40450790       3.77936555       2.69835174
C        6.78940790       8.44676555       3.86495174
C        9.46220790       0.29346555       1.22135174
C        6.86240790       4.66356555       1.31205174
C        9.38910790       4.07666555       3.77425174
C        5.62785940       8.14946555      10.04654794
C       10.33145940       0.59066555       5.25114794
C        5.70085940       4.96076555       7.49374794
C       10.25835940       3.77936555       7.80404794
C        6.64325940       8.44676555       8.97064794
C        9.31605940       0.29346555       6.32704794
C        6.71625940       4.66356555       6.41774794
C        9.24295940       4.07666555       8.87994794
H        5.18881980       7.27716555       4.60381594
H        0.13091980       1.46296555       0.49551594
H        5.26191980       5.83316555       2.05091594
H        0.05781980       2.90706555       3.04831594
H        0.99921980       7.80176555       0.72891594
H        4.32061980       0.93846555       4.37031594
H        0.92611980       5.30856555       3.28181594
H        4.39361980       3.43166555       1.81751594
H        2.91771980       7.72996555       3.02561594
H        2.40201980       1.01016555       2.07371594
H        2.99081980       5.38026555       0.47271594
H        2.32901980       3.35986555       4.62651594
H        5.04267130       7.27716555       9.70951214
H       -0.01522870       1.46296555       5.60121214
H        5.11577130       5.83316555       7.15661214
H       -0.08832870       2.90706555       8.15401214
H        0.85307130       7.80176555       5.83461214
H        4.17447130       0.93846555       9.47601214
H        0.77997130       5.30856555       8.38751214
H        4.24747130       3.43166555       6.92321214
H        2.77157130       7.72996555       8.13131214
H        2.25587130       1.01016555       7.17941214
H        2.84467130       5.38026555       5.57841214
H        2.18287130       3.35986555       9.73221214
H       10.65470790       7.27716555       4.59735174
H        5.59680790       1.46296555       0.48905174
H       10.72780790       5.83316555       2.04445174
H        5.52370790       2.90706555       3.04185174
H        6.46510790       7.80176555       0.72245174
H        9.78650790       0.93846555       4.36385174
H        6.39200790       5.30856555       3.27535174
H        9.85950790       3.43166555       1.81105174
H        8.38360790       7.72996555       3.01915174
H        7.86790790       1.01016555       2.06725174
H        8.45670790       5.38026555       0.46625174
H        7.79490790       3.35986555       4.62005174
H       10.50855940       7.27716555       9.70304794
H        5.45065940       1.46296555       5.59474794
H       10.58165940       5.83316555       7.15014794
H        5.37755940       2.90706555       8.14754794
H        6.31895940       7.80176555       5.82814794
H        9.64035940       0.93846555       9.46954794
H        6.24585940       5.30856555       8.38104794
H        9.71335940       3.43166555       6.91674794
H        8.23745940       7.72996555       8.12484794
H        7.72175940       1.01016555       7.17294794
H        8.31055940       5.38026555       5.57194794
H        7.64875940       3.35986555       9.72574794
O        1.27951980       0.70906555       3.14441594
O        4.04031980       8.03106555       1.95481594
O        1.35261980       3.66096555       0.59161594
O        3.96721980       5.07916555       4.50771594
O        2.25881980       7.51926555       3.77841594
O        3.06091980       1.22086555       1.32081594
O        2.33191980       5.59096555       1.22551594
O        2.98781980       3.14916555       3.87371594
O        1.13337130       0.70906555       8.25011214
O        3.89417130       8.03106555       7.06051214
O        1.20647130       3.66096555       5.69731214
O        3.82107130       5.07916555       9.61341214
O        2.11267130       7.51926555       8.88411214
O        2.91477130       1.22086555       6.42651214
O        2.18577130       5.59096555       6.33121214
O        2.84167130       3.14916555       8.97941214
O        6.74540790       0.70906555       3.13795174
O        9.50620790       8.03106555       1.94835174
O        6.81850790       3.66096555       0.58515174
O        9.43310790       5.07916555       4.50125174
O        7.72470790       7.51926555       3.77195174
O        8.52680790       1.22086555       1.31435174
O        7.79780790       5.59096555       1.21905174
O        8.45370790       3.14916555       3.86725174
O        6.59925940       0.70906555       8.24364794
O        9.36005940       8.03106555       7.05404794
O        6.67235940       3.66096555       5.69084794
O        9.28695940       5.07916555       9.60694794
O        7.57855940       7.51926555       8.87764794
O        8.38065940       1.22086555       6.42004794
O        7.65165940       5.59096555       6.32474794
O        8.30755940       3.14916555       8.97294794
\#Succinic acid molecule
14
Lattice="25.0 0.0 0.0 0.0 25.0 0.0 0.0 0.0 25.0"
H       14.11975000      15.46600000      11.93510000
H       12.78155000       9.53400000      13.87910000
H       12.43075000      11.33220000      10.96430000
H       10.77975000      11.35030000      11.58510000
H       11.26695000      13.64630000      10.87690000
H       11.27085000      13.66630000      12.64070000
C       11.79605000      11.73530000      11.76590000
C       11.79605000      13.26260000      11.76380000
C       13.20035000      13.84150000      11.73820000
C       12.27005000      11.15750000      13.08830000
O       14.22025000      13.22590000      11.50900000
O       13.18995000      15.18010000      11.98440000
O       12.41625000      11.77360000      14.12310000
O       12.49785000       9.81900000      12.99210000
\endgroup%
\newpage

\newpage
\clearpage
\bibliography{ref}
\bibliographystyle{ieeetr}

\end{document}